\documentclass[aps,prb,reprint,longbibliography,superscriptaddress]{revtex4-2}

\usepackage{graphicx,amsmath,amssymb,color,amsfonts,bm}
\usepackage[dvipsnames]{xcolor}

\usepackage[colorlinks=true]{hyperref}  
\hypersetup{
    bookmarks=true,        
    unicode=false,          
    pdftoolbar=true,     
    pdfmenubar=true,    
    pdffitwindow=false,  
    pdfstartview={FitH},  
    pdftitle={},   
    pdfauthor={},    
    pdfsubject={},  
    pdfcreator={},
    pdfproducer={}, 
    pdfkeywords={} {} {}, 
    pdfnewwindow=true,    
    colorlinks=true,     
    linkcolor=blue, 
    citecolor=blue,       
    filecolor=magenta,     
    urlcolor=blue,        
        breaklinks=true
} 

\newcommand{\be}{\begin{equation}}
\newcommand{\ee}{\end{equation}}
\newcommand{\bea}{\begin{eqnarray}}
\newcommand{\eea}{\end{eqnarray}}

\newcommand{\kv}[1]{{{\bm k}_{#1}}}

\begin{document}

\title{Far-from-equilibrium universality in two-dimensional Heisenberg antiferromagnets}

\author{Zhaoyi Li}
\affiliation{\small\it Department of Physics, Stanford University, Stanford, CA 94305, USA}

\author{Paolo Glorioso}
\affiliation{\small\it Department of Physics, Stanford University, Stanford, CA 94305, USA}

\author{Joaquin F. Rodriguez-Nieva}
\affiliation{\small\it Department of Physics, Stanford University, Stanford, CA 94305, USA}
\affiliation{\small\it Department of Physics, Texas A\&M University, College Station, TX 77843, USA}

\begin{abstract} 

We study the far-from-equilibrium dynamics of isolated two-dimensional Heisenberg antiferromagnets. We consider spin spiral initial conditions which imprint a position-dependent staggered-magnetization (or Neel order) in the two-dimensional lattice. Remarkably, we find a long-lived prethermal regime characterized by self-similar behavior of staggered magnetization fluctuations, although the system has no long-range order at finite energy and the staggered magnetization does not couple with conserved charges. Exploiting the separation of length scales introduced by the initial condition, we derive a simplified analytical model that allow us to compute the spatial-temporal scaling exponents and power-law distribution of the staggered magnetization fluctuations analytically, and find excellent agreement with numerical simulations using phase space methods. The scaling exponents are insensitive to details of the initial condition, in particular, no fine-tuning of energy is required to trigger the self-similar scaling regime. Compared with recent results on far-from-equilibrium universality on the Heisenberg ferromagnet, we find quantitatively distinct spatial-temporal scaling exponents, therefore suggesting that the same model with ferromagnetic and antiferromagnetic initial conditions can host different universal regimes. Our predictions are relevant to ultra-cold atoms simulators of Heisenberg magnets and driven antiferromagnetic insulators.

\end{abstract}

\maketitle

\section{Introduction}

Many-body systems out of thermodynamic equilibrium can exhibit universal phenomena beyond conventional equilibrium paradigms. Prominent examples include turbulence\cite{1999RMP_turbulence,bookzakharov,booknazarenko}, ageing\cite{2005ageingcalabrese}, coarsening\cite{1994Bray}, surface growth\cite{1986PRL_KPZ}, breakdown of transport\cite{forster1977large}, 
and percolation\cite{2014tauberbook}. One common theme in the study of equilibrium and out-of-equilibrium universality is the emergence of self-similar behavior: microscopically distinct models can be classified into universality classes sharing the same scaling exponents. Unlike systems at thermodynamic equilibrium, far-from-equilibrium systems break a symmetry associated to detailed-balance\cite{Sieberer2015,Crossley:2015evo,2018SciPost_neqdetailedbalance} and can therefore exhibit richer behaviors than their equilibrium counterparts with quantitatively distinct scaling exponents. Such rich behaviors have been observed in driven-dissipative systems\cite{2006PRL_mitra,2012PRL_noisydrivensystems,2013PRL_diehl,2016PRB_marino,2016PRB_Diehl} where an external drive pushes the system to a non-thermal steady state while dissipation maintains energy balance, and in quenches of isolated systems where the system acts as its own bath\cite{2002PRL_berges,2004berges,2008berges,2011PRD_berges,2013PRB_chandran,2014heavyions,2015berges,2015PRE_gambassi,2016PRB_mitra,2015bergesreview,2016PRA_tsubota,2017PRD_berges,2019PRL_gasenzer,2017PRL_Diehl}. These universal non-equilibrium regimes are now routinely probed in cold atom experiments\cite{2016Nature_hadzibabic,2018natureuniversality1,2018natureuniversality2,2018natureuniversality3,2021NP_hadzibabic}.

Conserved charges and order parameters, both of which are determined by symmetries and dimensionality of the system, play an important role in determining the nature of the scaling behavior. In some cases, non-conventional scaling behavior arises from the non-linear dynamics of conserved charges in phase space. Examples include turbulent phenomena in fluids and non-thermal fixed points of bosonic theories where the self-similar scaling is governed by the {\it cascade} of conserved charges in momentum space\cite{bookzakharov,2015bergesreview,2016PRA_tsubota,2017PRD_berges,2019PRL_gasenzer}, one-dimensional integrable systems where Kardar-Parisi-Zhang (KPZ) scaling arises due to the extensive number of conserved charges\cite{integrabilitybreaking3,2018PRLbulchandani,2020PRL_denardis,2019PRL_sarang}, and thermalizing systems in low dimensions which display a breakdown of local hydrodynamics \cite{forster1977large,das2020nonlinear,delacretaz2020breakdown} (see Ref.\cite{2022NP_glorioso} for an example in kinematically constrained systems). 
In other cases, non-conventional scaling is induced by the dynamics of the order parameter. Examples include coarsening dynamics where self-similar scaling arises due to the growth of ordered domains\cite{1994Bray,1995rutenberg,1995PRE_bray}, or ageing dynamics where self-similar scaling arises due to quasi-long range correlations close to criticality\cite{2014PRL_schmalian,2015PRB_schmalian,2013PRB_dynamicaltransitions,2017PRL_Diehl}. In general, quantities which are neither conserved nor couple with an order parameter are expected to relax in microscopic times and exhibit featureles fluctuations. 

\begin{figure}
\centering\includegraphics{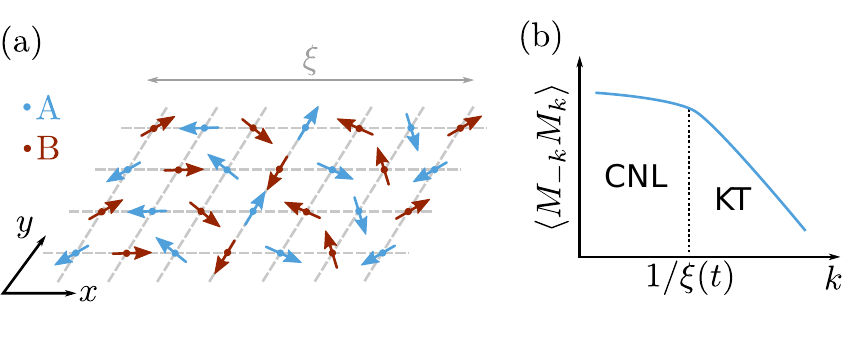}
\caption{(a) Schematics of the staggered spin spiral initial state, see Eq.(\ref{eq:initial}). The wavevector ${\bm q} = (q_x,0)$ of the initial conditions imprints a lengthscale $\xi(0)=1/q_x$ in the system at time $t=0$. (b) We compute the scaling behavior of $\xi(t)\sim t^\beta$ using a continuous non-linear (CNL) theory describing staggered magnetization fluctuations at small wavevectors $k\lesssim 1/\xi$, and we compute the power law distribution $\langle M_{-k}M_{k}\rangle\sim 1/k^\nu$ using a kinetic theory (KT) of magnons at wavevectors $k\gtrsim 1/\xi$. } 
\label{fig:schematics}
\end{figure}

Here we study the prethermal dynamics of isolated Heisenberg antiferromagnets and show that quantities which are neither conserved nor exhibit ordering at finite temperature can still exhibit slow relaxation and universal prethermal dynamics when initialized from certain excited states. In particular, using spin spiral excited states with inhomogeneous Neel order (see Fig.\ref{fig:schematics}), we find that large and slowly-relaxing staggered magnetization fluctuations persist during a long-lived prethermal regime although the system exhibits no symmetry-breaking phase transition at finite temperature. In this prethermal regime, the staggered magnetization rapidly relaxes to zero but staggered magnetization fluctuations exhibit universal scaling given by:
\be
C_{\bm k}(t)=\langle M_{-\bm k}(t) M_{\bm k}(t)\rangle = t^{\alpha}f(t^\beta|{\bm k}|),
\label{eq:scaling}
\ee
with $\beta \approx 0.5$ and $\alpha\approx 1$. In Eq.(\ref{eq:scaling}), the universal function $f$ also exhibits scaling $f(x) \sim 1/|x|^{\nu}$ for a broad range of momenta, with $\nu \approx 2.3$. The scaling exponents ($\alpha,\beta,\nu$) are universal in the sense that they are insensitive to details of the Hamiltonian or the initial condition. In particular, we emphasize that no fine-tuning of energy is needed. Using a continuous non-linear model describing long-wavelength spin modes combined with a kinetic theory of interacting quasiparticles at shorter wavelengths, we are able to derive analytically the three scaling exponents $(\alpha,\beta,\nu)$. These analytic values are shown to agree remarkably well with numerical simulations using phase space methods. 

We find that the origin of the prethermal scaling is associated to the existence of spin modes whose gapless nature is protected by the global SU(2) symmetry. We show this by computing the unequal time spin-spin correlation function\cite{2020_Onasier}, which exhibits linearly-dispersing gapless modes even though the prethermal state is far from the staggered ground state. A central insight in understanding the prethermal scaling in two-dimensional systems with SU(2) symmetry was discussed in recent works by one of us in the context of Heisenberg ferromagnets\cite{2020PRL_rodrigueznieva,2022PNAS_rodrigueznieva}. In the ferromagnetic case, the two-dimensional nature of the system bestows total magnetization fluctuations a long-range character. In addition, the combination of conserved magnetization and the constrained interactions resulting from the SU(2) symmetry gives rise to a universality class distinct from previously-studied instances of scaling. 
Similarly to the ferromagnetic case, the two-dimensional nature of the antiferromagnetic system can lead to quasi-long-range correlations of the staggered magnetization. However, unlike the ferromagnet, the staggered magnetization is not conserved. In spite of its non-conserved nature, we still find a parametrically long time window (controlled by non-universal parameters that depend on Hamiltonian details and the initial conditions) in which the system exhibits universal prethermal scaling. 

As discussed in more detail below, we find that the Heisenberg antiferromagnet belongs to a different non-equilibrium universality class than previously-studied models with similar features. Compared to the Heisenberg ferromagnet, we find clearly distinct scaling exponents in spite of the underlying Hamiltonian being the same up to an overall sign. We attribute such differences to the different dispersion of emergent gapless modes and the different effective interactions between such modes, see Sec.\,\ref{sec:analyt}. We also emphasize that the scaling regime discussed in the present work is intrinsically different from the predictions of model G in the Halperin-Hohenberg classification\cite{1977PR_HH}: while the latter describes universal behaviour close to thermodynamic equilibrium, here we consider a dynamical regime where equilibrium properties, such as the fluctuation-dissipation relation, are violated. In comparison with $O(n)$ theories in $d=2$, we note that these also exhibit linearly-dispersing quasiparticles and long-range order at $T=0$\cite{booksachdev}. We find that both $O(n)$ theories and Heisenberg antiferromagnets exhibit similar (within numerical uncertainty) non-equilibrium spatial-temporal scaling exponents $\beta\approx 0.5$ and $\alpha \approx 1$ while showing clearly distinct values of the universal scaling exponent $\nu$. The same conclusion applies when comparing non-relativistic bosonic theories and antiferromagnets (we note that relativistic $O(N)$ theories and non-relativistic $U(1)$ theories were shown to exhibit the same $\alpha = d\beta$ and $\beta\approx 1/2$ exponents\cite{2015asier}). We find that the difference in the exponent $\nu$ is related to the soft nature of the interaction between spin modes. We also note that the universal scaling in $O(n)$ and $U(n)$ theories is induced by gapped modes, whereas in antiferromagnets we observe gapless spin modes irrespective of the energy of the initial conditions.

Our predictions are relevant in a variety of experimental scenarios, both in condensed matter and cold atomic platforms. First, in recent years there has been remarkable advances in our experimental capabilities to probe the dynamics of isolated spin systems in low dimensions\cite{2014spiralstate,2020Nature_jepsen,2021PRX_jepsen}. Current experiments are now able to prepare simple product states of excited spin spirals and tune the dimensionality and exchange interactions to probe the dynamics under different symmetries. Such experiments are now able to coherently evolve the system over unprecedentedly long timescales on the order of $t \sim 50 \hbar/J$\cite{2020Nature_jepsen} and, therefore, access the long-lived prethermal regime discussed in the present work. On a different front, experiments in solid-state systems are now able to drive low-dimensional ferromagnetic and antiferromagnetic insulators and probe magnetization fluctuations with energy resolution using local probes\cite{2017chunhui,2018nv-ferro,2021pnas_tony,2021PRApplied_chunhui,2021PRL_chunhui}, such as nitrogen-vacancy centers in diamond. Such experiments are capable of 
directly measuring the power-law distribution in Eq.(\ref{eq:scaling}) as well as the relaxation dynamics of highly excited states. 

The outline of the manuscript is as follows: In Sec.\,\ref{sec:model}, we describe the physical model and the intitial conditions. In Sec.\,\ref{sec:analyt}, we present a simple statistical mechanics model used to derive the scaling exponents analytically. In Sec.\,\ref{sec:num}, we use phase space methods to numerically evaluate the scaling exponents. In Sec. \ref{sec:concl}, we discuss the connections between our work and previously-studied instances of scaling in related models, and also present the conclusions. In the appendices we provide additional details about asymptotic functions used in Sec.\,\ref{sec:analyt} (Appendix \ref{app:dm}) and details about the statistical analysis used to analyze the numerical data (Appendix \ref{app:num}). 

\section{Microscopic model}\label{sec:model}

We consider the Heisenberg antiferromagnet on a two-dimensional square lattice with nearest neighbour interactions:
\be
H = J\sum_{\langle ij \rangle} {S}_{i}^x {S}_{j}^x + {S}_{i}^y {S}_{j}^y +S_i^zS_j^z. 
\label{eq:Hafm}
\ee
Here ${\bm S}_i = (S_i^x,S_i^y,S_i^z)$ are spin-$S$ operators, $\langle ij \rangle$ denotes summation over nearest neighbor spins $i$ and $j$, and $J$ is a positive constant. For the purposes of our work, including additional next-nearest neighbour exchange does not affect the universal aspects of the dynamics, so long as the next-nearest-neighbor exchange does not induce frustration. As such, we restrict our discussion to nearest neighbour coupling only in order to keep the model as simple as possible. However, we emphasize that the SU(2) symmetry is essential to our discussion and breaking it will lead to qualitatively distinct results, as we discuss in more detail below. 

For a large spin number $S$ and zero temperature, the Heisenberg antiferromagent exhibits a broken-symmetry ground state with a non-zero staggered magnetization (or Neel order). In this broken-symmetry state, the staggered magnetization $M^a = \sum_i (-1)^{r_i^x+r_i^y}S_i^a$ has finite expectation value, with ${\bm r}_i = (r_i^x,r_i^y)$ the lattice coordinate of spin $i$ (expressed in units of lattice constant, thus, $r_{i}^{x,y}$ are integer numbers). Unlike the ferromagnetic case ($J<0$), the antiferromagnetic ground state is not classical and has zero point motion due to quantum fluctuations, {\it i.e.}, $\langle M^a \rangle_{T=0} < NS$. Interestingly, the broken-symmetry ground state is believed to be present even in the $S=1/2$ limit\cite{1988PRL_afmgroundstate,1988PRB_afmgroundstate}. For this reason, we expect that the universal aspects of dynamics discussed here will be independent of the spin number $S$, although part of our analysis relies on the classical limit $S\gg 1$. 

We consider an antiferromagnetic spin spiral product state as initial condition:
\be
\begin{array}{rl}
\langle S_{i}^{\pm}\rangle = & S(-1)^{r_i^x+r_i^y}\sin\theta e^{\pm i{\bm q}\cdot{\bm r}_i},\\
\langle S_i^z\rangle =& S(-1)^{r_i^x+r_i^y}\cos\theta,
\end{array}
\label{eq:initial}
\ee
where $S_i^\pm = S_i^x\pm S_i^y$. 
The initial condition (\ref{eq:initial}) imprints a lengthscale $\xi = 1/|{\bm q}|$ in which spins have Neel order (although there is no order globally)
while restricting dynamics in the zero magnetization sector $S^{\rm tot} = 0$. Dynamics of spin systems under this type of initial conditions are now routinely accessed in cold atomic platforms\cite{2014spiralstate,2020Nature_jepsen,2015PRX-babadi,2020PRBspinspiral}.

\section{Analytical derivation of the non-equilibrium scaling exponents}\label{sec:analyt}

In this section, we proceed to construct a simplified non-equilibrium statistical mechanics model that captures the universal aspects of prethermalization in  Heisenberg antiferromagnets, namely, the asymptotic form of the function $f(x) \sim 1/x^{\nu}$  and the numerical values of $(\alpha,\beta)$ in Eq.(\ref{eq:scaling}). The essence of the approach follows closely that used in Refs.\cite{2020PRB_spinturbulence,2022PNAS_rodrigueznieva} and relies on exploiting  the lengthscale separation of spin modes introduced by the initial condition, {\it i.e.}, $1 \ll \xi \ll L$.  The value of $\xi$, which is time-dependent, defines a correlation length for staggered magnetization fluctuations. There are two types of excitations that need to be incorporated into our description. First, the initial excited state will trigger spatial fluctuations of Neel order which govern the dynamics of the staggered magnetization at wavevectors $|{\bm k}| \lesssim 1/\xi$. Secondly, large-wavevector $|\bm k|\gtrsim 1/\xi$ excitations will mediate the transfer of energy and magnetization towards UV degrees of freedom. Whereas the former is described with a classical continuum theory, the latter is described using kinetic theory for magnons within the wave turbulence formalism. 

Following Ref.\cite{2020PRB_spinturbulence,2022PNAS_rodrigueznieva} we proceed in two steps. For wavevectors $|{\bm k}|\gtrsim 1/\xi$, we study the dynamics of magnetization fluctuations by assuming that the staggered magnetization is uniform in space: large wavevector magnons $|{\bm k}|\gtrsim 1/\xi$ effectively see a uniform magnetization background within the collision timescale. As such, in Sec.\,\ref{sec:magnons} we pin the antiferromagnetic order parameter and derive the effective kinetic theory for magnons. Using the wave turbulence formalism, we derive the power law exponent $\nu$ in Sec.\,\ref{sec:turbulence}. 
In Sec.\,\ref{sec:continuum}, we study dynamics of the order parameter within a continuum theory and determine the relevant non-linearities that govern the resulting spatial-temporal scaling of magnetization fluctuations for wavevectors $|\bm k|\lesssim 1/\xi$. From a simple scaling analysis that employs the equations of motion and spin conservation laws we derive the scaling exponents $(\alpha,\beta)$ that govern the growth of $\xi \sim t^\beta$.

\subsection{Dynamics of short-wavelengths}
\label{sec:magnons}

In the presence of a uniform staggered magnetization, it is convenient to write the spin operators in terms of bosonic operators. Following Ref.\cite{1971PRB_AFMhh}, we employ here the Dyson-Maleev transformation. Because of the antiferromagnetic nature of the ground state, we use two flavors of bosons to describe spins in the $A$ and $B$ sublattices (see Fig.\ref{fig:schematics}):
\begin{subequations}
\begin{align}
    &S_i^z = S - a_i^\dagger a_i,\label{eq:dysonmalev1} \\
    &S_i^+ = \sqrt{2S}\left(a_i - \frac{1}{2S}a_i^\dagger a_i a_i\right),\\
    &S_i^- = \sqrt{2S}a_i^\dagger,\\
    &S_j^z = -S + b_i^\dagger b_i, \\
    &S_j^+ = \sqrt{2S}\left(b_i - \frac{1}{2S}b_i^\dagger b_i b_i\right),\\
    &S_j^- = \sqrt{2S}b_i^\dagger.
    \label{eq:dysonmalev2}
\end{align}
\end{subequations}
Here $S_i^\pm$ denotes $S_i^\pm=S_i^x\pm i S_i^y$, $a_i$ and $b_j$ are bosonic annihilation operators, and we use $i$ ($j$) to label sites in the $A$ ($B$) sublattice. We note that the Dyson-Maleev transformation produces the correct spin commutation relations but violates the property $(S_i^-)^\dagger = S_i^+$. In this case, the spin Hamiltonian becomes non-Hermitian. In contrast, the Holstein-Primakoff transformation---another commonly used transormation for spin systems\cite{1940PR_holteinprimakoff}---retains Hermiticity but generates an infinite series of interaction vertices. In spite of the non-hermiticity property, the Dyson-Maleev transformation has proven to be more convenient for studying spin-wave interaction in ferromagnets and antiferromagnets: it reproduces all the perturbative results obtained with the Holstein-Primakoff transformation in a much faster and compact way\cite{1971PRB_AFMhh,1992PRB_afmdysonmaleev,1992PRB_afmdysonmaleev1}. 

Inserting Eqs.(\ref{eq:dysonmalev1})-(\ref{eq:dysonmalev2}) into Eq.\,(\ref{eq:Hafm}) and separating terms order by order, we find $H = -NJS^2 + H_0 + H_{\rm int}$, with
\begin{subequations}
\begin{align}
&H_0 = JS\sum_{\langle ij\rangle}\left( a_i^\dagger a_i + b_j^\dagger b_j + a_ib_j + b_j^\dagger a_i^\dagger\right),\\
&H_{\rm int} = -\frac{J}{2}\sum_{\langle ij\rangle}\left(a_i^\dagger a_i a_i b_j + a_i^\dagger b_j^\dagger b_j^\dagger b_j + 2a_i^\dagger a_i b_j^\dagger b_j\right). 
\end{align}
\label{eq:HDM}
\end{subequations}
Note that interactions are ${\cal O}(1/S)$ at small boson densities and vanish in the classical limit $S\rightarrow \infty$. Going into Fourier space and defining $a_{{\bm k}}=\frac{1}{\sqrt{N}}\sum_ie^{-i{\bm k}\cdot{\bm r}_i}a_i$, $b_{{\bm k}}=\frac{1}{\sqrt{N}}\sum_ie^{-i{\bm k}\cdot{\bm r}_j}b_j$, the quadratic component of the Hamiltonian is given by
\be
    H_0 = JzS\sum_{\bm k}(a^\dagger_{\bm k} a_{\bm k}+b^\dagger_{\bm k} b_{\bm k}+\gamma_{\bm k}a^\dagger_{\bm k}b^\dagger_{-{\bm k}}+\gamma_{\bm k}a_{\bm k}b_{-{\bm k}}).
    \label{eq:H0}
\ee
Here $\gamma_{\bm k}$ denotes the phase factor $\gamma_{\bm k}=\frac{1}{z} \sum_{\bm \ell}e^{i{\bm k}\cdot{\bm \ell}}$, where $\bm \ell$ denotes the nearest neighbor lattice vectors ${\bm\ell}=\{(\pm 1,0),(0,\pm 1)\}$, and $z$ is the coordination number of each spin ($z=4$ in a two-dimensional square lattice). The quartic component of the Hamiltonian is given by
\begin{align}
\nonumber H_{\rm int}=-&\frac{Jz}{2N}\sum_{\kv{1}\kv{2}\kv{3}\kv{4}}\delta(\kv{\rm i} - \kv{\rm f})\left( \gamma_{\kv{4}}a_{\kv{1}}^\dagger a_{-\kv{2}}a_{\kv{3}}b_{\kv{4}} \right. \\ \nonumber& \left.+\gamma_{\kv{1}}a_{\kv{1}}^\dagger b_{\kv{2}}^\dagger b_{-\kv{3}}^\dagger b_{\kv{4}}+ 2\gamma_{\kv{3}-\kv{2}}a_{\kv{1}}^\dagger b_{\kv{2}}^\dagger b_{\kv{3}}a_{\kv{4}}\right),
\end{align}
with ${\bm k}_{\rm i} ={\bm k}_1+{\bm k}_2$ and ${\bm k}_{\rm f}={\bm k}_3+{\bm k}_4$.

We now proceed to diagonalize $H_0$ using the Bogoliubov transformation
\begin{subequations}
\begin{align}
    a_{\bm k}= u_{\bm k}\alpha_{\bm k} - v_{\bm k}\beta_{-{\bm k}}^\dagger,\\
    b_{-{\bm k}}^\dagger = -v_{\bm k}\alpha_{\bm k}+u_{\bm k}\beta_{-{\bm k}}^\dagger,
\end{align}
\label{eq:bogoliubov}
\end{subequations}
where $\alpha_{\bm k}$ and $\beta_{\bm k}$ are bosonic annihilation operators (bosonic commutation requires $u_{\bm k}^2-v_{\bm k}^2 = 1$). Replacing Eq.(\ref{eq:bogoliubov}) into Eq.(\ref{eq:H0}) leads to 
\be
H_0 = \sum_{\bm k}\varepsilon_{\bm k}(\alpha_{\bm k}^\dagger \alpha_{\bm k} + \beta_{\bm k}^\dagger\beta_{\bm k}),\quad
\varepsilon_{\bm k} = zJS \sqrt{1-\gamma_{\bm k}},
\label{eq:H01}
\ee
and the factors $u_{\bm k}$ and $v_{\bm k}$ are given by
\be
u_{\bm k} = \sqrt{\frac{1+\varepsilon_{\bm k}}{2\varepsilon_{\bm k}}},\quad v_{\bm k} = \sqrt{\frac{1-\varepsilon_{\bm k}}{2\varepsilon_{\bm k}}}.
\label{eq:uv}
\ee
Using the Bogoliubov transformation on the quartic components of $H$ results in
\begin{widetext}
\begin{align}
    H_{\rm int}=-\frac{Jz}{2N}\sum_{{\kv{1}\kv{2}\kv{3}\kv{4}}}&\delta(\kv{\rm i}-\kv{\rm f})u_{\kv{1}}u_{\kv{2}}u_{\kv{3}}u_{\kv{4}}(\Phi^{(1)}_{\kv{1}\kv{2}\kv{3}\kv{4}}\alpha_{\kv{1}}^\dagger\alpha_{\kv{2}}^\dagger\alpha_{\kv{3}}\alpha_{\kv{4}} \nonumber\\
    &+2\Phi^{(2)}_{\kv{1}\kv{2}\kv{3}\kv{4}}\alpha_{\kv{1}}^\dagger\beta_{-{\kv{2}}}\alpha_{\kv{3}}\alpha_{\kv{4}}+2\Phi^{(3)}_{\kv{1}\kv{2}\kv{3}\kv{4}}\alpha_{\kv{1}}^\dagger\alpha_{\kv{2}}^\dagger\alpha_{\kv{3}}\beta_{-{\kv{4}}}^\dagger\nonumber\\  
        &+4\Phi^{(4)}_{{\kv{1}\kv{2}\kv{3}\kv{4}}}\alpha_{\kv{1}}^\dagger\beta_{-{\kv{2}}}\alpha_{\kv{3}}\beta_{-{\kv{4}}}^\dagger+2\Phi^{(5)}_{\kv{1}\kv{2}\kv{3}\kv{4}}\beta_{-\kv{1}}\beta_{-\kv{2}}\alpha_{\kv{3}}\beta_{-{\kv{4}}}^\dagger\nonumber\\
    &+2\Phi^{(6)}_{\kv{1}\kv{2}\kv{3}\kv{4}}\alpha_{\kv{1}}^\dagger\beta_{-\kv{2}}\beta_{-\kv{3}}^\dagger\beta_{-\kv{4}}^\dagger
    +\Phi^{(7)}_{\kv{1}\kv{2}\kv{3}\kv{4}}\alpha_{\kv{1}}^\dagger\alpha_{\kv{2}}^\dagger\beta_{-\kv{3}}^\dagger\beta_{-\kv{4}}^\dagger\nonumber\\
    &+\Phi^{(8)}_{\kv{1}\kv{2}\kv{3}\kv{4}}\beta_{-\kv{1}}\beta_{-\kv{2}}\alpha_{\kv{3}}\alpha_{\kv{4}}+\Phi^{(9)}_{\kv{1}\kv{2}\kv{3}\kv{4}}\beta_{-\kv{1}}\beta_{-\kv{2}}\beta_{-\kv{3}}^\dagger\beta_{-\kv{4}}^\dagger).\label{eq:Hint}
\end{align}
\end{widetext}
where the phase factors $\Phi_{\kv{1}\kv{2}\kv{3}\kv{4}}^{(n)}$ are explicitly written in Appendix \ref{app:dm}.

We now analyze Eqs.(\ref{eq:H0}) and (\ref{eq:Hint}) in the long wavelength limit in order to determine the power laws characterizing the quasiparticle dispersion and their interactions. From Eq.(\ref{eq:H0}) one finds the well-known dispersion of small-momenta magnons given by $\varepsilon_{\bm k}\propto |{\bm k}|$. A second useful relation that results from the linearized analysis is the ratio between the amplitude of the staggered magnetization and the total magnetization. This relation can be obtained directly from the Bogoliubov eigenvectors $(u_{\bm k},v_{\bm k})$ in Eq.(\ref{eq:uv}). Focusing on the staggered and total magnetization produced by the $\alpha_{\bm k}$ mode, we find that the staggered magnetization scales as $m_{\bm k} \approx a_{\bm k}-b_{\bm k}\approx (u_{\bm k}+v_{\bm k})\alpha_{\bm k}$, whereas the total magnetization scales as $s_{\bm k} = (a_{\bm k} + b_{\bm k})\approx (u_{\bm k}-v_{\bm k})\alpha_{\bm k}$. Using the small wavevector expansion of Eq.(\ref{eq:uv}), 
$u_{\bm k}\pm v_{\bm k} \approx \frac{1}{\sqrt{2\varepsilon_{\bm k}}}\left(1 \pm 1 +\frac{\varepsilon_{\bm k}\mp\varepsilon_{\bm k}}{2}\right) $, and $\varepsilon_{\bm k}\propto |{\bm k}|$ leads to $s_{\bm k} \approx |\bm k| m_{\bm k} \ll m_{\bm k}$, 
thus the total magnetization fluctuations are ${\cal O}(|{\bm k}|)$ smaller that the staggered magnetization fluctuations and vanish in the $|{\bm k}|\rightarrow 0$ limit, as expected for an antiferromagnet. 

We now proceed to analyze the scaling of $H_{\rm int}$ in Eq.(\ref{eq:Hint}) in the small wavevector limit. A detailed analysis of magnon relaxation in different energy ranges was done in Ref.\,\cite{1971PRB_AFMhh}. Of primary interest in this work are on-shell processes describing particle-conserving collision between magnons with wavevectors $|{\bm k}| \gtrsim 1/\xi$. Off-resonant processes and processes which do not preserve particle number, both of which we will neglect, were shown to give subleading effects in the relaxation dynamics at small wavevectors\cite{1971PRB_AFMhh}. Thus, the relevant terms in Eq.(\ref{eq:Hint}) are those containing the phase factors $\Phi^{(1)}$, $\Phi^{(4)}$, and $\Phi^{(9)}$, with $\Phi^{(1)} = \Phi^{(9)}$. For small wavevectors, the product $u_{\kv{1}}u_{\kv{2}}u_{\kv{3}}u_{\kv{4}}$ scales as $u_{\kv{1}}u_{\kv{2}}u_{\kv{3}}u_{\kv{4}}\sim \frac{1}{\sqrt{\varepsilon_{\kv{1}}\varepsilon_{\kv{2}}\varepsilon_{\kv{3}}\varepsilon_{\kv{4}}}}\sim 1/k^2$. In addition, the asymptotic form of $\Phi^{(1,9)}$ is given by $\Phi_{\kv{1}\kv{2}\kv{3}\kv{4}}^{(1,9)} = 2\varepsilon_{\kv{3}}\varepsilon_{\kv{4}}(\hat{\bm k}_3\cdot\hat{\bm k}_4-1)$ and the asymptotic form of $\Phi^{(4)}$ is given by $\Phi_{\kv{1}\kv{2}\kv{3}\kv{4}}^{(4)} = 2\varepsilon_{\kv{3}}\varepsilon_{\kv{4}}(\hat{\bm k}_3\cdot\hat{\bm k}_4+1)$, see Appendix \ref{app:dm}, and they all scale as $\Phi^{(n)}_{\kv{1}\kv{2}\kv{3}\kv{4}}\sim k^2$ in the long wavelength limit. As such, the matrix element for the two-body interaction $(\kv{1},\kv{2})\rightarrow(\kv{3},\kv{4})$ scales as $V(\lambda\kv{1},\lambda\kv{2},\lambda\kv{3},\lambda\kv{4}) = \lambda^0 V(\kv{1},\kv{2},\kv{3},\kv{4})$. 

As a side remark, we note that if SU(2) symmetry is broken, for example by adding anisotropic exchange, then the scaling with momentum of the interactions changes altogether. In particular, the phase factors $\Phi^{(n)}_{\kv{1}\kv{2}\kv{3}\kv{4}}$ become wavevector-independent and this would change the dynamical scaling laws characterizing the prethermal regime. 
In contrast, adding next-nearest neighbour exchange terms which preserve SU(2) symmetry will not alter the scaling behavior described in this work. 

\subsection{Scaling of the envelope function $f$}
\label{sec:turbulence}

We now calculate the power law scaling of the function $f(x)\sim 1/x^{\nu}$ using wave turbulence theory. Wave turbulence \cite{bookzakharov,booknazarenko} provides a framework for computing the scaling of two-point correlation functions in far-from-equilibrium regimes when the system exhibits a weak coupling limit. In our case, the weak coupling limit is controlled by the parameter $1/S$, see Eq.(\ref{eq:HDM}). The starting point in wave turbulence theory is to assume incoherent dynamics of the bosonic degrees of freedom $\alpha_{\bm k}$ and $\beta_{\bm k}$, {\it i.e.}, $\langle \alpha_{\bm k}\rangle = \langle \beta_{\bm k}\rangle = 0$, which is equivalent to assuming that transverse magnetization fluctuations (relative to the direction of the Neel order) are incoherent. This approximation is only valid when $|\bm k|\gtrsim 1/\xi$; in the regime $|\bm k|\lesssim 1/\xi$, instead, the Neel order will change its orientation, thus leading to finite expectation values for $\langle\alpha_{\bm k}\rangle$ and $\langle \beta_{\bm k}\rangle$ (this regime will be analyzed in the next section). 
Under this approximation, each flavor of magnons is characterized by its occupation number $\langle\alpha_{\bm k}^\dagger \alpha_{\bm k}\rangle = n_{\alpha,\bm k}$ and $\langle\beta_{\bm k}^\dagger \beta_{\bm k}\rangle = n_{\beta,\bm k}$. 
The standard procedure in wave turbulence consists of: (i) deriving a kinetic equation from Eqs.(\ref{eq:H0}) and (\ref{eq:Hint}) describing the time evolution of $n_{\bm k}$, (ii) proposing a solution of the form $n_{\bm k} \propto |k|^{-\nu}$ , and (iii) finding $\nu$ that gives rise to a steady-state solution (here we assume that $n_{\bm k,\alpha}=n_{\bm k,\beta}=n_{\bm k}$). 

Following the discussion in the previous section, we employ an effective theory describing magnon excitations that only includes on-shell terms which preserve both particle number and energy. The exponent $\nu$ can only depend on the power $\gamma$ of the quasiparticle dispersion $\varepsilon_{\bm k} \propto |\bm k|^\gamma$ , the power $\delta$ of the interaction, $V (\lambda\kv{1}, \lambda\kv{2}, \lambda\kv{3}, \lambda\kv{4} ) = \lambda^\delta V (\kv{1}, \kv{2}, \kv{3}, \kv{4})$, and the system's dimension $d$. Based on the results of the previous section, we have $\gamma = 2$ and $\delta = 0$. As shown in Refs.\cite{bookzakharov,booknazarenko}, there are two non-thermal solutions with scaling exponents
\be
\nu_N = d+\frac{2\delta-\gamma}{3}, \quad \nu_E =d + \frac{2\delta}{3}. 
\label{eq:exponents}
\ee
The solution with scaling exponent $\nu_N$ is associated to a flux of quasiparticles cascading towards small momenta (inverse cascade), whereas the solution with scaling exponent $\nu_E = 2$ is associated to a flux of energy cascading towards large momenta (direct cascade). The wavevector range where particle number is concentrated, $k_N = \frac{1}{N}\int \frac{d{\bm k}}{(2\pi)^2}|\bm k|n_{\bm k}$, is where the quasiparticle cascade occurs, whereas the wavevector range where energy is concentrated, $k_E=\frac{1}{E}\int \frac{d{\bm k}}{(2\pi)^2}|\bm k|n_{\bm k}\varepsilon_{\bm k}$, is where the energy cascade occurs. Because $k_N<k_E$, the inverse cascade characterizes scaling in the small wavevector regime whereas the energy cascade characterizes scaling at larger wavevectors. Using the values $d,\delta,\gamma$ specific to our system, we find $\nu_N = 4/3$ and $\nu_E = 2$. 

Finally, we note that the scaling exponents $\nu_{N}$ and $\nu_E$ characterize the distribution of the occupation numbers $n_{\alpha,\bm k}$ and $n_{\beta,\bm k}$ of the bosonic degrees of freedom rather than the spin degrees of freedom. Thus, the last step in our calculation is to transform back from bosonic operators into spin operators. We note that the spin operators are related to $(\alpha_{\bm k},\beta_{\bm k})$ through a Bogoliubov transformation, see Eq.(\ref{eq:bogoliubov}). As such, the scaling with  momentum of the spin-spin correlation function $\langle M_{-\bm k}^aM_{\bm k}^a\rangle$ is given by $\langle M_{-\bm k}^aM_{\bm k}^a\rangle \sim n_{\bm k}/k\sim 1/k^{\nu_{N,E}+1}$, thus $\nu = \nu_{N,E}+1$. 

We comment on the validity of the exponents $\nu_N$ and $\nu_E$. In a strict sense, we need to question the validity of using a kinetic theory involving only on-resonant and particle conserving terms while neglecting all other processes. The validity of these approximations will be confirmed numerically in the next section. However, we emphasize that Ref.\,\cite{1971PRB_AFMhh} made a strong case to justify neglecting such processes for the smallest momenta magnons, which is the wavevector range described by the exponent $\nu_N$.  Instead, for larger-momenta magnons, which is the wavevector range described by the exponent $\nu_E$, off-resonant processes and processes which do not preserve particle number become incrementally more relevant and the kinetic approximation begins to break down. Indeed, in our numerics below we find that the scaling exponent of $f(x)\sim 1/x^\nu$ matches remarkably well with $\nu = \nu_N +1 $ but we do not observe a second wavevector range with exponent $\nu = \nu_E+1$, which is consistent with the picture presented in Ref.\cite{1971PRB_AFMhh}. 

\subsection{Spatio-temporal scaling exponents}
\label{sec:continuum}

In this section, we focus on the dynamics of magnetization fluctuations with wavevectors $|{\bm k}|\lesssim 1/\xi$. We use the microscopic equations of motion in the continuum limit and analyze the role of leading order non-linearities to phenomenologically predict the exponents $\alpha$ and $\beta$ governing the growth of $\xi(t)$. We define two spin fields $\bm a = \langle {\bm S}_{i\in A}\rangle$ and $\bm b = \langle {\bm S}_{j\in B}\rangle$ that characterize the average orientation of the spin $\bm S$ in sublattice A and B, respectively. We also define the staggered magnetization ${\bm m} = {\bm a}-{\bm b}$ and the total magnetization ${\bm s} = {\bm a}+{\bm b}$. Assuming that the fields ${\bm a}$ and $\bm b$ vary smoothly in space, we can expand the microscopic equations of motion for spins, $\partial_t {\bm S}_i = \sum_j {\bm S}_i\times {\bm S}_j$ to leading order in gradients:
\bea
\partial_t{\bm a} = {\bm a}\times(z{\bm b}+\nabla^2{\bm b}), \\ \partial_t{\bm b} = {\bm b}\times(z{\bm a}+\nabla^2{\bm a}), 
\eea
where time is expressed in units of $1/J$, and we approximated the sum over neighbouring spins with the laplacian, $\sum_{\bm\ell}{\bm a}_{{\bm x} + {\bm\ell}} \approx z{\bm b}_{\bm x}+\nabla^2{\bm b}_{\bm x}$ and $\sum_{\bm \ell}{\bm a}_{{\bm x}+{\bm\ell}} \approx z{\bm a}_{\bm x}+\nabla^2{\bm a}_{\bm x}$. In terms of $\bm m$ and $\bm s$, we find
\begin{subequations}
\begin{align}
\partial_t{\bm s} =& \frac{1}{2}{\bm s}\times\nabla^2{\bm s} - \frac{1}{2}{\bm m}\times\nabla^2{\bm m},\\
\partial_t{\bm m} =& \frac{z}{2}{\bm m}\times{\bm s} - \frac{1}{2}{\bm s}\times\nabla^2{\bm m} + \frac{1}{2}{\bm m}\times\nabla^2{\bm s}.
\label{eq:dmt}
\end{align}
\label{seq:eomminus}
\end{subequations}
To find the values of $(\alpha,\beta)$ in Eq.(\ref{eq:scaling}), following Ref.\cite{2022PNAS_rodrigueznieva} we look at Eq.(\ref{eq:dmt}) and balance the time derivative of $\bm m$ with the leading nonlinearity, which is given by the first term on the right-hand-side:
\be
\partial_t m_{\bm k}^a \approx \frac{z}{2}\epsilon_{abc} \sum_{\bm p}m_{\bm p-\bm k}^bs_{\bm p}^c,
\label{eq:eom}
\ee
with $\epsilon_{abc}$ the Levi-Civita symbol. 
The second approximation that we use is that $s_{\bm k}^a \sim |{\bm k}| m_{\bm k}^a$, which is justified from the linearized analysis of spin waves discussed above which showed that the total magnetization is ${\cal O}(k)$ smaller than the staggered magnetization. The final approximation that we use is to assume that all modes with wavevector $|\bm k|\lesssim 1/\xi$ are macroscopic and democratically occupied for all spin orientations. As such, if
we identify $\xi \sim t^\beta$ in Eq.(\ref{eq:scaling}) and take $f\sim {\cal O}(1)$ at small wavevectors, then $m_{\bm k}^a$ scales as $m_{\bm k}^a \sim \xi^{\alpha/2\beta}$ and $s_{\bm k}^a$ scales as $s_{\bm k}^a\sim \xi^{\alpha/2\beta-1}$. Using this scaling form in the left-hand-side of Eq.(\ref{eq:eom}), we find $\partial_t m_{\bm k}^a \sim \xi^{(\alpha-2)/2\beta} $, where we used $\dot\xi = \xi^{1-1/\beta}$. The right-hand side of Eq.(\ref{eq:eom}), instead, yields $\sum_{\bm k}m_{\bm p-\bm k}^b s_{\bm p}^c\sim \xi^{\alpha/\beta-d-1}$, where we approximated $\sum_{\bm k}=A\int \frac{d^d{\bm k}}{(2\pi)^d}\propto \xi^{-d}$ and $A$ is the total area of the system. Equating both sides of Eq.(\ref{eq:eom}) such that they both yield the same temporal scaling for $\xi(t)$ results in
\be
2(d+1)\beta = \alpha +2.
\label{seq:condition1}
\ee

The second relation between $\alpha$ and $\beta$ comes from the conservation of spin length, $\frac{1}{4}\int d{\bm x}({\bm m}+{\bm s})^2 + ({\bm m}-{\bm s})^2=$\,constant. Using the scaling $m_{\bm k}^a\sim \xi^{\alpha/2\beta}$ and neglecting the contribution of the total magnetization $|{\bm s}|\ll {\bm m}$, we find the second condition
\be
\alpha = d\beta.
\label{seq:condition2}
\ee
Combining Eq.(\ref{seq:condition1}) with (\ref{seq:condition2}) in $d=2$ yields
\be
\alpha = 1,\quad \beta = 1/2.
\ee
We note that a similar analysis in the ferromagnetic case resulted in quantitatively different exponents $\alpha = 2/3$ and $\beta = 1/3$\cite{2022PNAS_rodrigueznieva}, indicating that the Heisenberg ferromagnet and antiferromagnet belong to different non-equilibrium universality classes. In particular, the same analysis leading to Eq.(\ref{seq:condition1}) resulted in the condition  $2(d+2)\beta = \alpha+2$. 

\section{Numerical simulations through phase space methods}\label{sec:num}

We compute the real time dynamics of quantum spins using the Truncated Wigner Approximation (TWA)\cite{2010review_polkovnikov,spintwa1,spintwa2,spintwa3}.  This method incorporates quantum fluctuations by adding quantum noise in the initial conditions and evolving the classical trajectories using the classical equations of motion for spins $\partial_{t}{\bm S}_{i}=J\sum_{j}{\bm S}_{i}\times {\bm S}_{j}$. 
To sample the initial conditions in Eq.(\ref{eq:initial}), we use a Gaussian approximation for the Wigner function given by 
\be
W(S^\perp_i,S^z_i)=\frac{2}{\pi S}e^{\frac{-(S_i^\perp)^2}{S}}\delta{(S_i^z-S)}, 
\label{eq:twasampling}
\ee
which reproduces the correct first and second moment of the Wigner distribution\cite{2010review_polkovnikov}. In Eq.(\ref{eq:twasampling}), we assumed without loss of generality that the initial spin is pointing in the $+z$ direction. 

\begin{figure}
\centering\includegraphics{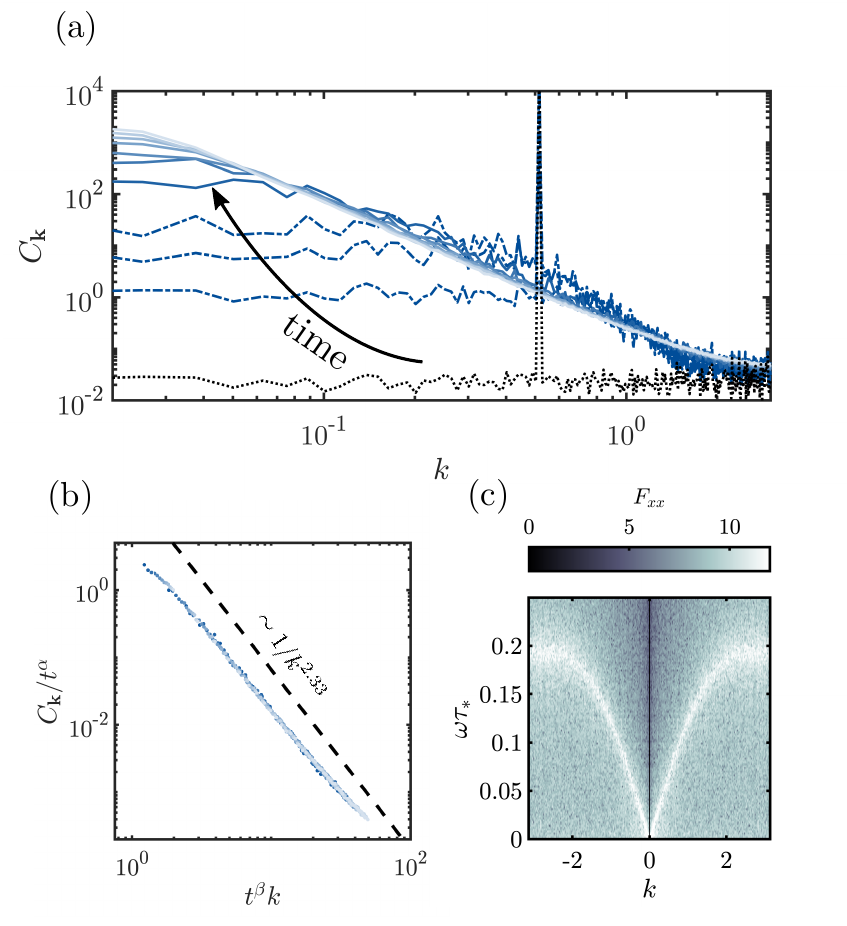}
\caption{(a) Evolution of the spin-spin correlation function for the staggered magnetization, $C_{\bm k}(t)=\sum_{a}\langle {M}^a_{\bf k}{M}^a_{-\bf k}\rangle$. Shown with dotted lines is the correlation function for the initial state, with dashed-dotted lines is the correlation function prior to the self-similar regime, and with solid lines is the correlation function in the self-similar scaling regime. Lighter shade of color indicate increasing times. (b) Re-scaled spin-spin correlation function using Eq.(\ref{eq:scaling}), with $\alpha=1$ and $\beta=0.5$. The dashed line indicates the power law scaling $\sim x^{-7/3}$. (c) Unequal time spin-spin correlation function exhibiting a linearly-dispersing gapless modes at small momenta. 
Simulation parameters: $L = 500, q_x = 0.5, \theta=\frac{\pi}{2}, S =10$.}
\label{fig:selfsimilar}
\end{figure}

Figure \ref{fig:selfsimilar} shows the time evolution of the equal-time spin-spin correlation function for the staggered magnetization $C_{\bm k}(t) = \sum_{a=x,y,z}{\langle{M}^a_{\bm k}(t){M}^a_{-\bm k}(t)\rangle}$ for a system of linear size $L=500$, and initial conditions with wavevector $q_x = 0.5$, $q_y=0$, and $\theta = \pi/2$. At $t=0$, only a single mode with wavevector ${\bm k}=(q_x,0)$ is macroscopically occupied. Within a timescale on the order of the inverse energy (per spin), the macroscopic state is depleted and a power law distribution of the two-point correlation function develops. In this prethermal regime, we find that the values
\be
\alpha = 1.0\pm 0.1 ,\quad \beta = 0.48 \pm0.05,
\label{eq:alphabeta}
\ee
best fit the numerical data in a sufficiently long time window, see details in Appendix \ref{app:num}. These values agree with the analytical predictions of the previous section using scaling arguments. 

We now proceed to analyze the power law scaling of the function $f$ in Eq.(\ref{eq:scaling}), see dashed lines in Fig.\ref{fig:selfsimilar}(b). Interestingly, we observe only {\it one} power law characterizing the tails 
of the magnetization fluctuation distribution, contrary to the predictions of wave turbulence found in Sec.\,\ref{sec:turbulence} which suggested the existence of two exponents $\nu_N$ and $\nu_E$. In fitting the power law exponent of the distribution we find that 
\be
\nu = 2.4\pm 0.1, 
\label{eq:nu}
\ee
which agrees with the exponent $\nu = \nu_N+1 = 7/3\approx 2.33$ associated to the inverse particle cascade. This result is in agreement with the conclusions of Ref.\cite{1971PRB_AFMhh} which argues that off-resonant processes and particle-non-conserving can be neglected at small momenta, thus the effective theory (\ref{eq:H0})-(\ref{eq:Hint}) considering only particle conserving processes is a good approximation. For larger wavevectors (or energies), however, it is likely that off-resonant processes play a more prominent role and, therefore, the energy cascade exponent $\nu_E$ is not present. 

We tested the robustness of our results using initial conditions with different values of $\theta$, ranging from $\frac{\pi}{6}<\theta<\frac{\pi}{2} $ and different values of the wavevectors $q_x$, and we consistently see the same scaling exponents within
numerical uncertainty. Similarly to the Heisenberg ferromagnet, this suggests that the far-from-equilibrium dynamics of the isotropic Heisenberg antiferromagnet is governed by a {\it single} non-thermal fixed point with the exponents in (\ref{eq:alphabeta}) and (\ref{eq:nu}). In contrast, U(1) theories were shown to exhibit multiple non-thermal fixed points, each of which can be activated by different initial conditions. 

Whereas the groundstate of the Heisenberg antiferromagnet exhibits gapless low-energy excitations, it is unclear whether the highly excited initial condition in Eq.(\ref{eq:initial}) leads to a dynamically-generated gap, such as those observed in $O(n)$ and $U(n)$ theories, or whether the spin modes in the self-similar region remain gapless. In a recent work\cite{2022PNAS_rodrigueznieva}, it was shown that the global SU(2) symmetry of the Heisenberg ferromagnet precludes the opening of a dynamical gap during evolution, leading to a long-lived prethermal regime governed by gapless modes with dispersion $\omega_{\bm k} \sim |{\bm k}|^2$. We numerically checked the nature of the excitations at the lowest wavevectors using the unequal-time correlation function $F_{xx}({\bm k},\omega) = \int dt e^{i\omega t}\langle\frac{1}{2}\{M_{-\bm k}^x(t_0+t)M_{\bm k}^x(t_0)\} \rangle $ for the staggered magnetization ($\{A,B\} = AB+BA$), see Fig.\ref{fig:selfsimilar}(c). Interestingly, we observe that the self-similar regime is governed by gapless modes at all times, even when the intermediate-time prethermal state is far from the ground state with uniform Neel order. The dispersion of the gapless mode is consistent with $\omega_{\bm k} \sim k$, different from the $\omega_{\bm k}\sim k^2$ dispersion observed in the Heisenberg ferromagnet. 

\section{Discussion and summary}\label{sec:concl}

The scaling regime discussed in the present work is intrinsically different from previously-studied instances of scaling in several important ways. Compared to the universal prethermal dynamics of  Heisenberg ferromagnets\cite{2022PNAS_rodrigueznieva}, we find clearly distinct exponents originating from the existence of gapless modes with linear dispersion rather than modes with quadratic dispersion. In addition, unlike the ferromagnetic case, antiferromagnetic fluctuations are not coupled to any conserved charge. As such, the long-lived prethermal regime discussed here is cut off by some parametrically long timescale controlled by processes that give rise to staggered magnetization decay, see particle non-conserving terms in Eq.\,(\ref{eq:Hint}). 

Heisenberg antiferromagnets and $O(n)$ theories share many similarities at thermodynamic equilibrium and low temperatures, in particular, both exhibit linearly-dispersing quasiparticles and a symmetry-breaking phase transition at $T=0$ ($T>0$) in dimension $d=2$ ($d>2$). We also find that they share some similarities in far-from-equilibrium regimes. For example, several works found the same scaling exponents $\alpha = d\beta$ and $\beta \approx 1/2$ in $O(n)$ theories regardless of the value of $n$\cite{2015asier,2020_Onasier}. However, in $O(n)$ theories quenched to (or across) a critical point \cite{2013PRB_chandran,2015PRE_gambassi,2017PRL_Diehl} self-similarity occurs only if parameters and initial conditions are fine-tuned so as to guarantee a vanishing late-time effective gap, unlike the antiferromagnetic case where we observe scaling without any fine-tuning of the initial conditions. In addition, we observe quantitative differences in the universal scaling function $f$ between both theories, reinforcing the idea that both models do not belong to the same non-equilibrium universality class. 

Compared to non-thermal fixed points in bosonic $U(1)$ theories, we note that models with $U(1)$ symmetry in two dimensions can exhibit topological defects (vortices) which can qualitatively alter the far-from-equilibrium behavior and give rise to different self-similar scaling regimes\cite{2011PRB_gasenzer,2012PRA_gasenzer}. Even in the absence of vortices, the effective theory for the antiferromagnet (see Sec. \ref{sec:analyt}) differs from the $U(1)$ bosonic theory both at the level of quasiparticle dispersion and their effective interactions, suggesting that both cannot belong to the same universality class. In certain cases, an effective gap has been observed to be dynamically generated by fluctuations \cite{2015asier,2020_Onasier}. This effective gap has
been shown to lead to a modified non-relativistic effective theory at low momenta and share the same scaling exponents $\alpha = d\beta$ and $\beta \approx 1/2$ characterizing $O(n)$ theories and Heisenberg antiferromagnets. However, at the level of the universal scaling function $f$ in Eq.(\ref{eq:scaling}) we find a clearly distinct exponent $\nu$ which sets the dynamics of 
antiferromagnets and $U(1)$ theories apart. 

In summary, we studied the universal far-from-equilibrium dynamics of two-dimensional Heisenberg antiferromagnets. We showed that, if initialized in a state with inhomogeneous Neel order, magnetization fluctuations will exhibit a long-lived prethermal regime with universal behavior. This shows that quantities which are neither conserved nor exhibit long range order can still exhibit self-similiar behavior in a parametrically long time window. The scaling exponents are shown to be remarkably robust to details of the initial conditions---in particular, no fine-tuning of the energy is necessary. Our work also highlights the important role played by dimensionality and symmetry in giving rise to gapless spin modes with long range character. Combined with a recent work by one of us on Heisenberg ferromagnets\cite{2022PNAS_rodrigueznieva}, we have now fully characterized the non-thermal fixed points of the Heisenberg model both for ferromagnetic and antiferromagnetic exchange. The scaling regime discussed in this work is readily accessible in ongoing experiments in cold atomic gases which can probe these regimes in fully-tunable spin systems\cite{2020Nature_jepsen,2003amospinexchange,2019amospin}, including tunable symmetries and spatial dimension.

\section*{Acknowledgements} 

We are grateful to Jamir Marino and Asier Pi\~{n}eiro-Orioli for insightful comments and previous collaborations. JFRN acknowledges the Gordon and Betty Moore Foundation’s EPiQS Initiative through Grant GBMF4302 and GBMF8686, the 2021 KITP program {\it Non-equilibrium universality: from classical to quantum and back}, and the National Science Foundation under Grant No. NSF PHY-1748958. PG is supported by the Alfred P. Sloan Foundation through Grant FG-2020-13615, the Department of Energy through Award DE-SC0019380, and the Simons Foundation through Award No. 620869.        

\appendix

\renewcommand{\thefigure}{A\arabic{figure}}
\setcounter{figure}{0}

\section{Interaction coefficients within the Dyson-Maleev formalism}\label{app:dm}

Here we reproduce the phase factor coefficients of the interactions in the Heisenberg Hamiltonian [Eq.(\ref{eq:Hint})] within the Dyson-Maleev transformation [\ref{eq:dysonmalev1}] using the notation in Ref.\cite{1971PRB_AFMhh}. In particular, the phase factors $\Phi^{(n)}$ appearing in the interactions of the effective Hamiltonian (\ref{eq:Hint}) are given by 
\begin{subequations}
\begin{align}
    \nonumber&\Phi^{(1,9)}_{\kv{1}\kv{2}\kv{3}\kv{4}}=(
     \gamma_{{\kv{1}}-{\kv{4}}}x_{\kv{1}}x_{\kv{4}}
    +\gamma_{{\kv{1}}-{\kv{3}}}x_{\kv{1}}x_{\kv{3}}\\\nonumber
    +&\gamma_{\kv{2}-{\kv{4}}}x_\kv{2}x_{\kv{4}}
    +\gamma_{\kv{2}-{\kv{3}}}x_\kv{2}x_{\kv{3}}
    -\gamma_{\kv{1}}x_\kv{2}x_{\kv{3}}x_{\kv{4}}\\
    -&\gamma_\kv{2}x_{\kv{1}}x_{\kv{3}}x_{\kv{4}}
    -\gamma_\kv{2}x_\kv{2}
    -\gamma_{\kv{1}}x_{\kv{1}}),\\\nonumber
    &\Phi^{(2,4)}_{\kv{1}\kv{2}\kv{3}\kv{4}}=(
    -\gamma_{\kv{2}-{\kv{4}}}x_{\kv{4}}
    -\gamma_{\kv{2}-{\kv{3}}}x_{\kv{3}}\\\nonumber
    -&\gamma_{{\kv{1}}-{\kv{4}}}x_{\kv{1}}x_\kv{2}x_{\kv{4}}
    -\gamma_{{\kv{1}}-{\kv{3}}}x_{\kv{1}}x_\kv{2}x_{\kv{3}}
    +\gamma_{\kv{1}}x_{\kv{3}}x_{\kv{4}}\\
    +&\gamma_\kv{2}x_{\kv{1}}x_\kv{2}x_{\kv{3}}x_{\kv{4}}
    +\gamma_\kv{2}
    +\gamma_{\kv{1}}x_{\kv{1}}x_\kv{2}),\\\nonumber
    &\Phi^{(3)}_{\kv{1}\kv{2}\kv{3}\kv{4}}=\Phi^{(5)}_{\kv{1}\kv{2}\kv{3}\kv{4}}=(
    -\gamma_{\kv{2}-{\kv{4}}}x_\kv{2}
    -\gamma_{{\kv{1}}-{\kv{4}}}x_{\kv{1}}\\\nonumber
    -&\gamma_{\kv{2}-{\kv{4}}}x_{\kv{1}}x_{\kv{3}}x_{\kv{4}}
    -\gamma_{\kv{2}-{\kv{3}}}x_\kv{2}x_{\kv{3}}x_{\kv{4}}
    +\gamma_{\kv{1}}x_\kv{2}x_{\kv{3}}\\
    +&\gamma_\kv{2}x_{\kv{1}}x_{\kv{3}}
    +\gamma_\kv{2}x_\kv{2}x_{\kv{4}}
    +\gamma_{\kv{1}}x_{\kv{1}}x_{\kv{4}}),\\\nonumber
    &\Phi^{(4)}_{\kv{1}\kv{2}\kv{3}\kv{4}}=(\gamma_{{\kv{2}}-{\kv{4}}}
    +\gamma_{{\kv{1}}-{\kv{4}}}x_{\kv{1}}x_\kv{2}
    +\gamma_{{\kv{1}}-{\kv{4}}}x_{\kv{3}}x_{\kv{4}}\\\nonumber
    +&\gamma_{{\kv{1}}-{\kv{3}}}x_{\kv{1}}x_\kv{2}x_{\kv{3}}x_{\kv{4}}
    -\gamma_{\kv{1}}x_{\kv{3}}
    -\gamma_\kv{2}x_{\kv{1}}x_\kv{2}x_{\kv{3}})\\
    -&\gamma_\kv{2}x_{\kv{4}}
    -\gamma_{\kv{1}}x_{\kv{1}}x_\kv{2}x_{\kv{4}},\\\nonumber
    &\Phi^{(7,8)}_{\kv{1}\kv{2}\kv{3}\kv{4}}=(
    \gamma_{\kv{2}-{\kv{4}}}x_\kv{2}x_{\kv{3}}
    +\gamma_{\kv{2}-{\kv{3}}}x_\kv{2}x_{\kv{4}}\\\nonumber
    +&\gamma_{\kv{2}-{\kv{3}}}x_{\kv{1}}x_{\kv{3}}
    +\gamma_{\kv{2}-{\kv{3}}}x_{\kv{1}}x_{\kv{3}}
    +\gamma_{\kv{2}-{\kv{4}}}x_{\kv{1}}x_{\kv{4}}\\
    -&\gamma_{\kv{1}}x_{\kv{1}}x_{\kv{3}}x_{\kv{4}}
    -\gamma_\kv{2}x_\kv{2}x_{\kv{3}}x_{\kv{4}}-\gamma_{\kv{1}}x_\kv{2}-\gamma_\kv{2}x_{\kv{1}}).
\end{align}
\label{eq:Phi}
 \end{subequations}
In these expressions, the parameter $\gamma_{\bm k}$ is given by 
$\gamma_{\bm k}=\frac{1}{z}\sum_{\bm \ell}e^{i{\bm k}\cdot{\bm \ell}}$ defined in the main text, where $\bm \ell$ is the nearest neighbour lattice vectors in two-dimensions. The parameter $x_{\bm k}$ is the ratio $x_{\bm k} = u_{\bm k}/v_{\bm k} = \sqrt{(1-\varepsilon_{\bm k})/(1+\varepsilon_{\bm k})}$, with $\varepsilon_{\bm k} = \sqrt{1-\gamma_{\bm k}^2}$. 

Of primary interest are the expressions in Eq.(\ref{eq:Phi}) in the long wavelength limit $|{\bm k}|\rightarrow 0$, particularly for the factors $\Phi^{(1)} = \Phi^{(9)}$, and $\Phi^{(4)}$ which contribute to particle-conserving scattering processes in the kinetic theory. We first note the following identities which hold in the asymptotic limit $|{\bm k}|\rightarrow0$:
\begin{subequations}
\begin{align}
    \varepsilon_{\bm k}&\approx\frac{1}{2}|{\bm k}|, \\
    \gamma_{\bm k}&\approx 1-\frac{1}{2}\varepsilon_{\bm k}^2\\
    x_{\bm k}&\approx 1-\varepsilon_{\bm k}.
\end{align}
\end{subequations}
Replacing these asymptotic expressions into Eq.(\ref{eq:Phi}) and taking $\kv{1}+\kv{2}=\kv{3}+\kv{4}$ due to momentum conservation leads to
\begin{subequations}
\begin{align}
    &\Phi^{(1,9)}_{\kv{1}\kv{2}\kv{3}\kv{4}}=\frac{1}{2}\kv{3}\cdot\kv{4}-2\varepsilon_{\kv{3}}\varepsilon_{\kv{4}},\\
    &\Phi^{(4)}_{\kv{1}\kv{2}\kv{3}\kv{4}}=\frac{1}{2}\kv{3}\cdot\kv{4}+2\varepsilon_{\kv{3}}\varepsilon_{\kv{4}},
\end{align}
\end{subequations}
to leading order in momentum $\bm k$. These relations can also be written as:
\begin{subequations}
\begin{align}
    \frac{1}{2}\Phi^{(1,9)}_{\kv{1}\kv{2}\kv{3}\kv{4}}=&\varepsilon_{\kv{3}}\varepsilon_{\kv{4}}(\hat{k}_3\cdot\hat{k}_4-1),\\\nonumber
    \frac{1}{2}\Phi^{(4)}_{\kv{1}\kv{2}\kv{3}\kv{4}}=&\varepsilon_{\kv{3}}\varepsilon_{\kv{4}}(\hat{k}_3\cdot\hat{k}_4+1).
\end{align}
 \end{subequations}
Importantly, these phase factors accounting for 
particle-conserving processes scale as $\Phi\sim {\bm k}^2$, which is used in the derivation of $\nu$ in Eq.(\ref{eq:nu}) of the main text.

\section{Statistical analysis of numerical data}\label{app:num}

\begin{figure}[b]
\centering\includegraphics[scale=0.5]{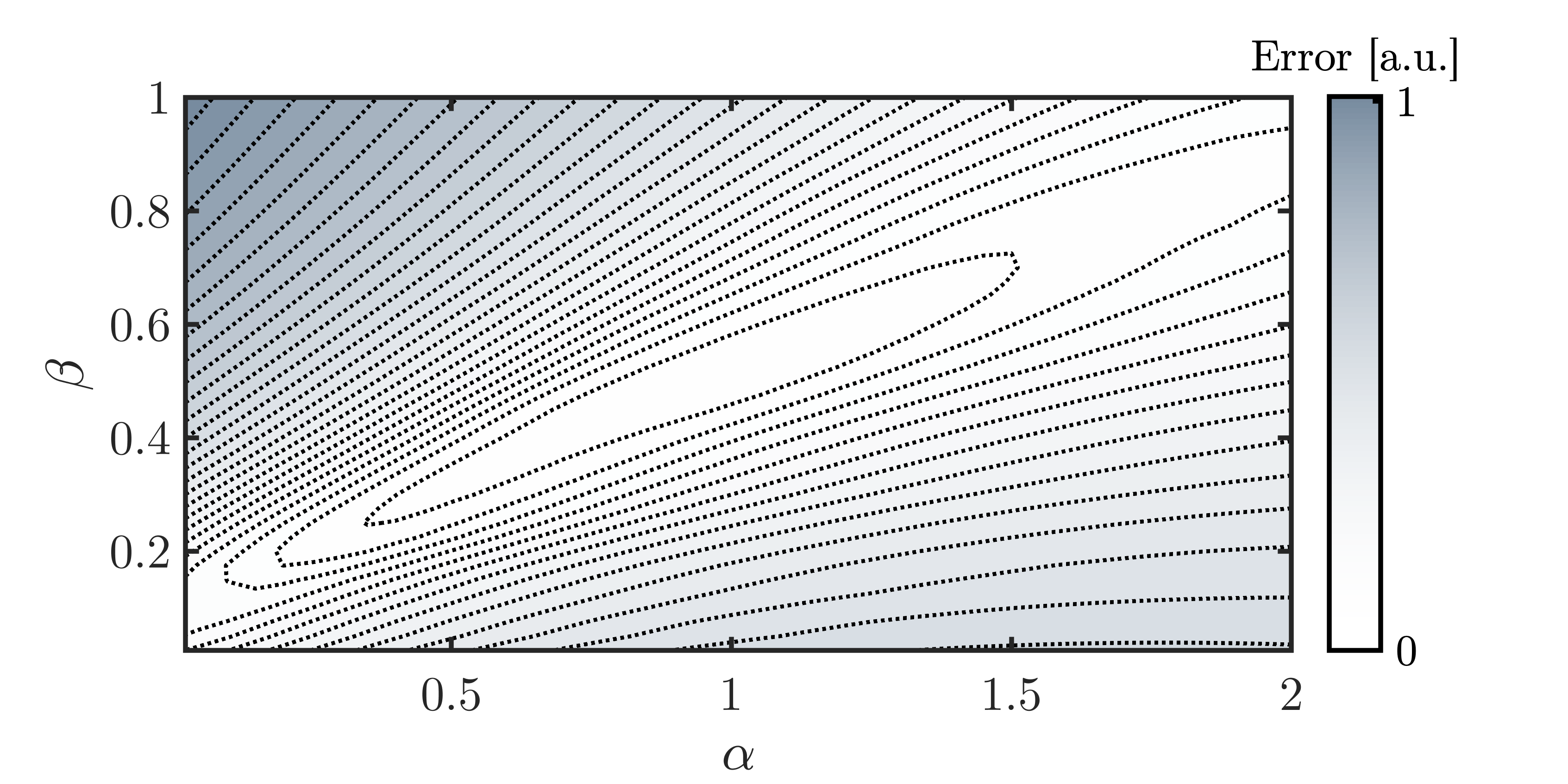}
\caption{Error function in Eq.(\ref{eq:error}) plotted as a function of the fitting parameters $\alpha$ and $\beta$. Darker color indicates higher error values. The minimum of $E$ occurs at around $\beta\approx 0.5$ and alpha $\alpha\approx 1$.}
\label{fig:error_contour}
\end{figure}

We obtain the scaling exponents $\alpha$ and $\beta$ that best fit the numerical data by minimizing the error function $E(\alpha,\beta)$ that quantifies the collapse of the data points through the scaling in Eq.(\ref{eq:scaling}). First, we take discrete values of $|\bm k|=k_i$ compatible with the inverse lattice spacing and evaluate the distribution $C(k,t_m)$ at different time steps $t_m$ within the self-similar regime ($t_{m+1}-t_m \sim \tau$ is roughly the inverse energy of the system). Second, we define the re-scaled variables $y_{i,m}=t^\alpha_mC(k_i,t_m)$ and $x_{i,m}=t^\beta_mk_i$. By interpolating these variable, we are able to obtain an explicit function $y_m(x)$, where $x$ is assumed to be a continuum variable. Third, we compute the error function as
\begin{equation}
    E(\alpha,\beta)=\sum_{m,m^\prime}\int dx|y_m(x)-y_{m^\prime}(x)|.
    \label{eq:error}
\end{equation}
The contour plot of $E(\alpha,\beta)$ for the self-similar regime studied in Fig.\ref{fig:selfsimilar} is shown in Figure~\ref{fig:error_contour}. We find that the best fitting with minimum error is centered around the point $(\alpha,\beta)\approx (1,0.5)$, which is consistent with our theoretical prediction. The error bars are obtained from the sensitivity of the parameter $(\alpha,\beta)$ for different initial conditions. 

\begin{thebibliography}{78}%
\makeatletter
\providecommand \@ifxundefined [1]{%
 \@ifx{#1\undefined}
}%
\providecommand \@ifnum [1]{%
 \ifnum #1\expandafter \@firstoftwo
 \else \expandafter \@secondoftwo
 \fi
}%
\providecommand \@ifx [1]{%
 \ifx #1\expandafter \@firstoftwo
 \else \expandafter \@secondoftwo
 \fi
}%
\providecommand \natexlab [1]{#1}%
\providecommand \enquote  [1]{``#1''}%
\providecommand \bibnamefont  [1]{#1}%
\providecommand \bibfnamefont [1]{#1}%
\providecommand \citenamefont [1]{#1}%
\providecommand \href@noop [0]{\@secondoftwo}%
\providecommand \href [0]{\begingroup \@sanitize@url \@href}%
\providecommand \@href[1]{\@@startlink{#1}\@@href}%
\providecommand \@@href[1]{\endgroup#1\@@endlink}%
\providecommand \@sanitize@url [0]{\catcode `\\12\catcode `\$12\catcode
  `\&12\catcode `\#12\catcode `\^12\catcode `\_12\catcode `\%12\relax}%
\providecommand \@@startlink[1]{}%
\providecommand \@@endlink[0]{}%
\providecommand \url  [0]{\begingroup\@sanitize@url \@url }%
\providecommand \@url [1]{\endgroup\@href {#1}{\urlprefix }}%
\providecommand \urlprefix  [0]{URL }%
\providecommand \Eprint [0]{\href }%
\providecommand \doibase [0]{http://dx.doi.org/}%
\providecommand \selectlanguage [0]{\@gobble}%
\providecommand \bibinfo  [0]{\@secondoftwo}%
\providecommand \bibfield  [0]{\@secondoftwo}%
\providecommand \translation [1]{[#1]}%
\providecommand \BibitemOpen [0]{}%
\providecommand \bibitemStop [0]{}%
\providecommand \bibitemNoStop [0]{.\EOS\space}%
\providecommand \EOS [0]{\spacefactor3000\relax}%
\providecommand \BibitemShut  [1]{\csname bibitem#1\endcsname}%
\let\auto@bib@innerbib\@empty
\bibitem [{\citenamefont {Sreenivasan}(1999)}]{1999RMP_turbulence}%
  \BibitemOpen
  \bibfield  {author} {\bibinfo {author} {\bibfnamefont {Katepalli~R.}\
  \bibnamefont {Sreenivasan}},\ }\bibfield  {title} {\enquote {\bibinfo {title}
  {Fluid turbulence},}\ }\href {\doibase 10.1103/RevModPhys.71.S383} {\bibfield
   {journal} {\bibinfo  {journal} {Rev. Mod. Phys.}\ }\textbf {\bibinfo
  {volume} {71}},\ \bibinfo {pages} {S383--S395} (\bibinfo {year}
  {1999})}\BibitemShut {NoStop}%
\bibitem{bookzakharov} V.E. Zakharov, V.S. L'vov, G. Falkovich, {\emph{Kolmogorov Spectra of Turbulence I: Wave Turbulence}}, Springer Berlin Heidelberg (1992)
\bibitem [{\citenamefont {Nazarenko}(2011)}]{booknazarenko}%
  \BibitemOpen
  \bibfield  {author} {\bibinfo {author} {\bibfnamefont {S.}~\bibnamefont
  {Nazarenko}},\ }\href {https://books.google.com/books?id=cAHZ2aEMqgYC} {\emph
  {\bibinfo {title} {Wave Turbulence}}},\ Lecture Notes in Physics\ (\bibinfo
  {publisher} {Springer Berlin Heidelberg},\ \bibinfo {year}
  {2011})\BibitemShut {NoStop}%
\bibitem [{\citenamefont {Calabrese}\ and\ \citenamefont
  {Gambassi}(2005)}]{2005ageingcalabrese}%
  \BibitemOpen
  \bibfield  {author} {\bibinfo {author} {\bibfnamefont {Pasquale}\
  \bibnamefont {Calabrese}}\ and\ \bibinfo {author} {\bibfnamefont {Andrea}\
  \bibnamefont {Gambassi}},\ }\bibfield  {title} {\enquote {\bibinfo {title}
  {Ageing properties of critical systems},}\ }\href {\doibase
  10.1088/0305-4470/38/18/r01} {\bibfield  {journal} {\bibinfo  {journal}
  {Journal of Physics A: Mathematical and General}\ }\textbf {\bibinfo {volume}
  {38}},\ \bibinfo {pages} {R133--R193} (\bibinfo {year} {2005})}\BibitemShut
  {NoStop}%
\bibitem [{\citenamefont {Bray}(1994)}]{1994Bray}%
  \BibitemOpen
  \bibfield  {author} {\bibinfo {author} {\bibfnamefont {A.J.}\ \bibnamefont
  {Bray}},\ }\bibfield  {title} {\enquote {\bibinfo {title} {Theory of
  phase-ordering kinetics},}\ }\href {\doibase 10.1080/00018739400101505}
  {\bibfield  {journal} {\bibinfo  {journal} {Advances in Physics}\ }\textbf
  {\bibinfo {volume} {43}},\ \bibinfo {pages} {357--459} (\bibinfo {year}
  {1994})}\BibitemShut {NoStop}%
\bibitem [{\citenamefont {Kardar}\ \emph {et~al.}(1986)\citenamefont {Kardar},
  \citenamefont {Parisi},\ and\ \citenamefont {Zhang}}]{1986PRL_KPZ}%
  \BibitemOpen
  \bibfield  {author} {\bibinfo {author} {\bibfnamefont {Mehran}\ \bibnamefont
  {Kardar}}, \bibinfo {author} {\bibfnamefont {Giorgio}\ \bibnamefont
  {Parisi}}, \ and\ \bibinfo {author} {\bibfnamefont {Yi-Cheng}\ \bibnamefont
  {Zhang}},\ }\bibfield  {title} {\enquote {\bibinfo {title} {Dynamic scaling
  of growing interfaces},}\ }\href {\doibase 10.1103/PhysRevLett.56.889}
  {\bibfield  {journal} {\bibinfo  {journal} {Phys. Rev. Lett.}\ }\textbf
  {\bibinfo {volume} {56}},\ \bibinfo {pages} {889--892} (\bibinfo {year}
  {1986})}\BibitemShut {NoStop}%
\bibitem [{\citenamefont {Forster}\ \emph {et~al.}(1977)\citenamefont
  {Forster}, \citenamefont {Nelson},\ and\ \citenamefont
  {Stephen}}]{forster1977large}%
  \BibitemOpen
  \bibfield  {author} {\bibinfo {author} {\bibfnamefont {Dieter}\ \bibnamefont
  {Forster}}, \bibinfo {author} {\bibfnamefont {David~R}\ \bibnamefont
  {Nelson}}, \ and\ \bibinfo {author} {\bibfnamefont {Michael~J}\ \bibnamefont
  {Stephen}},\ }\bibfield  {title} {\enquote {\bibinfo {title} {Large-distance
  and long-time properties of a randomly stirred fluid},}\ }\href@noop {}
  {\bibfield  {journal} {\bibinfo  {journal} {Physical Review A}\ }\textbf
  {\bibinfo {volume} {16}},\ \bibinfo {pages} {732} (\bibinfo {year}
  {1977})}\BibitemShut {NoStop}%
\bibitem [{\citenamefont {T{\"a}uber}(2014)}]{2014tauberbook}%
  \BibitemOpen
  \bibfield  {author} {\bibinfo {author} {\bibfnamefont {Uwe~C}\ \bibnamefont
  {T{\"a}uber}},\ }\href@noop {} {\emph {\bibinfo {title} {Critical dynamics: a
  field theory approach to equilibrium and non-equilibrium scaling behavior}}}\
  (\bibinfo  {publisher} {Cambridge University Press},\ \bibinfo {year}
  {2014})\BibitemShut {NoStop}%
\bibitem [{\citenamefont {Sieberer}\ \emph {et~al.}(2015)\citenamefont
  {Sieberer}, \citenamefont {Chiocchetta}, \citenamefont {Gambassi},
  \citenamefont {T{\"a}uber},\ and\ \citenamefont {Diehl}}]{Sieberer2015}%
  \BibitemOpen
  \bibfield  {author} {\bibinfo {author} {\bibfnamefont {LM}~\bibnamefont
  {Sieberer}}, \bibinfo {author} {\bibfnamefont {A}~\bibnamefont
  {Chiocchetta}}, \bibinfo {author} {\bibfnamefont {Andrea}\ \bibnamefont
  {Gambassi}}, \bibinfo {author} {\bibfnamefont {UC}~\bibnamefont
  {T{\"a}uber}}, \ and\ \bibinfo {author} {\bibfnamefont {S}~\bibnamefont
  {Diehl}},\ }\bibfield  {title} {\enquote {\bibinfo {title} {Thermodynamic
  equilibrium as a symmetry of the {Schwinger}-{Keldysh} action},}\ }\href@noop
  {} {\bibfield  {journal} {\bibinfo  {journal} {Physical Review B}\ }\textbf
  {\bibinfo {volume} {92}},\ \bibinfo {pages} {134307} (\bibinfo {year}
  {2015})}\BibitemShut {NoStop}%
\bibitem [{\citenamefont {Crossley}\ \emph {et~al.}(2017)\citenamefont
  {Crossley}, \citenamefont {Glorioso},\ and\ \citenamefont
  {Liu}}]{Crossley:2015evo}%
  \BibitemOpen
  \bibfield  {author} {\bibinfo {author} {\bibfnamefont {Michael}\ \bibnamefont
  {Crossley}}, \bibinfo {author} {\bibfnamefont {Paolo}\ \bibnamefont
  {Glorioso}}, \ and\ \bibinfo {author} {\bibfnamefont {Hong}\ \bibnamefont
  {Liu}},\ }\bibfield  {title} {\enquote {\bibinfo {title} {{Effective field
  theory of dissipative fluids}},}\ }\href {\doibase 10.1007/JHEP09(2017)095}
  {\bibfield  {journal} {\bibinfo  {journal} {JHEP}\ }\textbf {\bibinfo
  {volume} {09}},\ \bibinfo {pages} {095} (\bibinfo {year} {2017})}\BibitemShut
  {NoStop}%
\bibitem [{\citenamefont {Aron}\ \emph {et~al.}(2018)\citenamefont {Aron},
  \citenamefont {Biroli},\ and\ \citenamefont
  {Cugliandolo}}]{2018SciPost_neqdetailedbalance}%
  \BibitemOpen
  \bibfield  {author} {\bibinfo {author} {\bibfnamefont {Camille}\ \bibnamefont
  {Aron}}, \bibinfo {author} {\bibfnamefont {Giulio}\ \bibnamefont {Biroli}}, \
  and\ \bibinfo {author} {\bibfnamefont {Leticia~F.}\ \bibnamefont
  {Cugliandolo}},\ }\bibfield  {title} {\enquote {\bibinfo {title} {{(Non)
  equilibrium dynamics: a (broken) symmetry of the Keldysh generating
  functional}},}\ }\href {\doibase 10.21468/SciPostPhys.4.1.008} {\bibfield
  {journal} {\bibinfo  {journal} {SciPost Phys.}\ }\textbf {\bibinfo {volume}
  {4}},\ \bibinfo {pages} {008} (\bibinfo {year} {2018})}\BibitemShut {NoStop}%
\bibitem [{\citenamefont {Mitra}\ \emph {et~al.}(2006)\citenamefont {Mitra},
  \citenamefont {Takei}, \citenamefont {Kim},\ and\ \citenamefont
  {Millis}}]{2006PRL_mitra}%
  \BibitemOpen
  \bibfield  {author} {\bibinfo {author} {\bibfnamefont {Aditi}\ \bibnamefont
  {Mitra}}, \bibinfo {author} {\bibfnamefont {So}~\bibnamefont {Takei}},
  \bibinfo {author} {\bibfnamefont {Yong~Baek}\ \bibnamefont {Kim}}, \ and\
  \bibinfo {author} {\bibfnamefont {A.~J.}\ \bibnamefont {Millis}},\ }\bibfield
   {title} {\enquote {\bibinfo {title} {Nonequilibrium quantum criticality in
  open electronic systems},}\ }\href {\doibase 10.1103/PhysRevLett.97.236808}
  {\bibfield  {journal} {\bibinfo  {journal} {Phys. Rev. Lett.}\ }\textbf
  {\bibinfo {volume} {97}},\ \bibinfo {pages} {236808} (\bibinfo {year}
  {2006})}\BibitemShut {NoStop}%
\bibitem [{\citenamefont {Dalla~Torre}\ \emph {et~al.}(2012)\citenamefont
  {Dalla~Torre}, \citenamefont {Demler}, \citenamefont {Giamarchi},\ and\
  \citenamefont {Altman}}]{2012PRL_noisydrivensystems}%
  \BibitemOpen
  \bibfield  {author} {\bibinfo {author} {\bibfnamefont {Emanuele~G}\
  \bibnamefont {Dalla~Torre}}, \bibinfo {author} {\bibfnamefont {Eugene}\
  \bibnamefont {Demler}}, \bibinfo {author} {\bibfnamefont {Thierry}\
  \bibnamefont {Giamarchi}}, \ and\ \bibinfo {author} {\bibfnamefont {Ehud}\
  \bibnamefont {Altman}},\ }\bibfield  {title} {\enquote {\bibinfo {title}
  {Dynamics and universality in noise-driven dissipative systems},}\
  }\href@noop {} {\bibfield  {journal} {\bibinfo  {journal} {Physical Review
  B}\ }\textbf {\bibinfo {volume} {85}},\ \bibinfo {pages} {184302} (\bibinfo
  {year} {2012})}\BibitemShut {NoStop}%
\bibitem [{\citenamefont {Sieberer}\ \emph {et~al.}(2013)\citenamefont
  {Sieberer}, \citenamefont {Huber}, \citenamefont {Altman},\ and\
  \citenamefont {Diehl}}]{2013PRL_diehl}%
  \BibitemOpen
  \bibfield  {author} {\bibinfo {author} {\bibfnamefont {LM}~\bibnamefont
  {Sieberer}}, \bibinfo {author} {\bibfnamefont {Sebastian~D}\ \bibnamefont
  {Huber}}, \bibinfo {author} {\bibfnamefont {E}~\bibnamefont {Altman}}, \ and\
  \bibinfo {author} {\bibfnamefont {S}~\bibnamefont {Diehl}},\ }\bibfield
  {title} {\enquote {\bibinfo {title} {Dynamical critical phenomena in
  driven-dissipative systems},}\ }\href@noop {} {\bibfield  {journal} {\bibinfo
   {journal} {Physical review letters}\ }\textbf {\bibinfo {volume} {110}},\
  \bibinfo {pages} {195301} (\bibinfo {year} {2013})}\BibitemShut {NoStop}%
\bibitem [{\citenamefont {Marino}\ and\ \citenamefont
  {Diehl}(2016)}]{2016PRB_marino}%
  \BibitemOpen
  \bibfield  {author} {\bibinfo {author} {\bibfnamefont {Jamir}\ \bibnamefont
  {Marino}}\ and\ \bibinfo {author} {\bibfnamefont {Sebastian}\ \bibnamefont
  {Diehl}},\ }\bibfield  {title} {\enquote {\bibinfo {title} {Quantum dynamical
  field theory for nonequilibrium phase transitions in driven open systems},}\
  }\href {\doibase 10.1103/PhysRevB.94.085150} {\bibfield  {journal} {\bibinfo
  {journal} {Phys. Rev. B}\ }\textbf {\bibinfo {volume} {94}},\ \bibinfo
  {pages} {085150} (\bibinfo {year} {2016})}\BibitemShut {NoStop}%
\bibitem [{\citenamefont {Chiocchetta}\ \emph
  {et~al.}(2016{\natexlab{a}})\citenamefont {Chiocchetta}, \citenamefont
  {Gambassi}, \citenamefont {Diehl},\ and\ \citenamefont
  {Marino}}]{2016PRB_Diehl}%
  \BibitemOpen
  \bibfield  {author} {\bibinfo {author} {\bibfnamefont {Alessio}\ \bibnamefont
  {Chiocchetta}}, \bibinfo {author} {\bibfnamefont {Andrea}\ \bibnamefont
  {Gambassi}}, \bibinfo {author} {\bibfnamefont {Sebastian}\ \bibnamefont
  {Diehl}}, \ and\ \bibinfo {author} {\bibfnamefont {Jamir}\ \bibnamefont
  {Marino}},\ }\bibfield  {title} {\enquote {\bibinfo {title} {Universal
  short-time dynamics: Boundary functional renormalization group for a
  temperature quench},}\ }\href@noop {} {\bibfield  {journal} {\bibinfo
  {journal} {Physical Review B}\ }\textbf {\bibinfo {volume} {94}},\ \bibinfo
  {pages} {174301} (\bibinfo {year} {2016}{\natexlab{a}})}\BibitemShut
  {NoStop}%
\bibitem [{\citenamefont {Aarts}\ and\ \citenamefont
  {Berges}(2002)}]{2002PRL_berges}%
  \BibitemOpen
  \bibfield  {author} {\bibinfo {author} {\bibfnamefont {Gert}\ \bibnamefont
  {Aarts}}\ and\ \bibinfo {author} {\bibfnamefont {Juergen}\ \bibnamefont
  {Berges}},\ }\bibfield  {title} {\enquote {\bibinfo {title} {Classical
  aspects of quantum fields far from equilibrium},}\ }\href {\doibase
  10.1103/PhysRevLett.88.041603} {\bibfield  {journal} {\bibinfo  {journal}
  {Phys. Rev. Lett.}\ }\textbf {\bibinfo {volume} {88}},\ \bibinfo {pages}
  {041603} (\bibinfo {year} {2002})}\BibitemShut {NoStop}%
\bibitem [{\citenamefont {Berges}\ \emph {et~al.}(2004)\citenamefont {Berges},
  \citenamefont {Bors\'anyi},\ and\ \citenamefont {Wetterich}}]{2004berges}%
  \BibitemOpen
  \bibfield  {author} {\bibinfo {author} {\bibfnamefont {J.}~\bibnamefont
  {Berges}}, \bibinfo {author} {\bibfnamefont {Sz.}\ \bibnamefont
  {Bors\'anyi}}, \ and\ \bibinfo {author} {\bibfnamefont {C.}~\bibnamefont
  {Wetterich}},\ }\bibfield  {title} {\enquote {\bibinfo {title}
  {Prethermalization},}\ }\href {\doibase 10.1103/PhysRevLett.93.142002}
  {\bibfield  {journal} {\bibinfo  {journal} {Phys. Rev. Lett.}\ }\textbf
  {\bibinfo {volume} {93}},\ \bibinfo {pages} {142002} (\bibinfo {year}
  {2004})}\BibitemShut {NoStop}%
\bibitem [{\citenamefont {Berges}\ \emph {et~al.}(2008)\citenamefont {Berges},
  \citenamefont {Rothkopf},\ and\ \citenamefont {Schmidt}}]{2008berges}%
  \BibitemOpen
  \bibfield  {author} {\bibinfo {author} {\bibfnamefont {J\"urgen}\
  \bibnamefont {Berges}}, \bibinfo {author} {\bibfnamefont {Alexander}\
  \bibnamefont {Rothkopf}}, \ and\ \bibinfo {author} {\bibfnamefont {Jonas}\
  \bibnamefont {Schmidt}},\ }\bibfield  {title} {\enquote {\bibinfo {title}
  {Nonthermal fixed points: Effective weak coupling for strongly correlated
  systems far from equilibrium},}\ }\href {\doibase
  10.1103/PhysRevLett.101.041603} {\bibfield  {journal} {\bibinfo  {journal}
  {Phys. Rev. Lett.}\ }\textbf {\bibinfo {volume} {101}},\ \bibinfo {pages}
  {041603} (\bibinfo {year} {2008})}\BibitemShut {NoStop}%
\bibitem [{\citenamefont {Berges}\ and\ \citenamefont
  {Sexty}(2011)}]{2011PRD_berges}%
  \BibitemOpen
  \bibfield  {author} {\bibinfo {author} {\bibfnamefont {J\"urgen}\
  \bibnamefont {Berges}}\ and\ \bibinfo {author} {\bibfnamefont {D\'enes}\
  \bibnamefont {Sexty}},\ }\bibfield  {title} {\enquote {\bibinfo {title}
  {Strong versus weak wave-turbulence in relativistic field theory},}\ }\href
  {\doibase 10.1103/PhysRevD.83.085004} {\bibfield  {journal} {\bibinfo
  {journal} {Phys. Rev. D}\ }\textbf {\bibinfo {volume} {83}},\ \bibinfo
  {pages} {085004} (\bibinfo {year} {2011})}\BibitemShut {NoStop}%
\bibitem [{\citenamefont {Chandran}\ \emph {et~al.}(2013)\citenamefont
  {Chandran}, \citenamefont {Nanduri}, \citenamefont {Gubser},\ and\
  \citenamefont {Sondhi}}]{2013PRB_chandran}%
  \BibitemOpen
  \bibfield  {author} {\bibinfo {author} {\bibfnamefont {Anushya}\ \bibnamefont
  {Chandran}}, \bibinfo {author} {\bibfnamefont {Arun}\ \bibnamefont
  {Nanduri}}, \bibinfo {author} {\bibfnamefont {Steven~S}\ \bibnamefont
  {Gubser}}, \ and\ \bibinfo {author} {\bibfnamefont {Shivaji~L}\ \bibnamefont
  {Sondhi}},\ }\bibfield  {title} {\enquote {\bibinfo {title} {Equilibration
  and coarsening in the quantum {$O(N)$} model at infinite {$N$}},}\
  }\href@noop {} {\bibfield  {journal} {\bibinfo  {journal} {Physical Review
  B}\ }\textbf {\bibinfo {volume} {88}},\ \bibinfo {pages} {024306} (\bibinfo
  {year} {2013})}\BibitemShut {NoStop}%
\bibitem [{\citenamefont {Berges}\ \emph {et~al.}(2014)\citenamefont {Berges},
  \citenamefont {Boguslavski}, \citenamefont {Schlichting},\ and\ \citenamefont
  {Venugopalan}}]{2014heavyions}%
  \BibitemOpen
  \bibfield  {author} {\bibinfo {author} {\bibfnamefont {J.}~\bibnamefont
  {Berges}}, \bibinfo {author} {\bibfnamefont {K.}~\bibnamefont {Boguslavski}},
  \bibinfo {author} {\bibfnamefont {S.}~\bibnamefont {Schlichting}}, \ and\
  \bibinfo {author} {\bibfnamefont {R.}~\bibnamefont {Venugopalan}},\
  }\bibfield  {title} {\enquote {\bibinfo {title} {Turbulent thermalization
  process in heavy-ion collisions at ultrarelativistic energies},}\ }\href
  {\doibase 10.1103/PhysRevD.89.074011} {\bibfield  {journal} {\bibinfo
  {journal} {Phys. Rev. D}\ }\textbf {\bibinfo {volume} {89}},\ \bibinfo
  {pages} {074011} (\bibinfo {year} {2014})}\BibitemShut {NoStop}%
\bibitem [{\citenamefont {Berges}\ \emph {et~al.}(2015)\citenamefont {Berges},
  \citenamefont {Boguslavski}, \citenamefont {Schlichting},\ and\ \citenamefont
  {Venugopalan}}]{2015berges}%
  \BibitemOpen
  \bibfield  {author} {\bibinfo {author} {\bibfnamefont {J.}~\bibnamefont
  {Berges}}, \bibinfo {author} {\bibfnamefont {K.}~\bibnamefont {Boguslavski}},
  \bibinfo {author} {\bibfnamefont {S.}~\bibnamefont {Schlichting}}, \ and\
  \bibinfo {author} {\bibfnamefont {R.}~\bibnamefont {Venugopalan}},\
  }\bibfield  {title} {\enquote {\bibinfo {title} {Universality far from
  equilibrium: From superfluid {Bose} gases to heavy-ion collisions},}\ }\href
  {\doibase 10.1103/PhysRevLett.114.061601} {\bibfield  {journal} {\bibinfo
  {journal} {Phys. Rev. Lett.}\ }\textbf {\bibinfo {volume} {114}},\ \bibinfo
  {pages} {061601} (\bibinfo {year} {2015})}\BibitemShut {NoStop}%
\bibitem [{\citenamefont {Maraga}\ \emph {et~al.}(2015)\citenamefont {Maraga},
  \citenamefont {Chiocchetta}, \citenamefont {Mitra},\ and\ \citenamefont
  {Gambassi}}]{2015PRE_gambassi}%
  \BibitemOpen
  \bibfield  {author} {\bibinfo {author} {\bibfnamefont {Anna}\ \bibnamefont
  {Maraga}}, \bibinfo {author} {\bibfnamefont {Alessio}\ \bibnamefont
  {Chiocchetta}}, \bibinfo {author} {\bibfnamefont {Aditi}\ \bibnamefont
  {Mitra}}, \ and\ \bibinfo {author} {\bibfnamefont {Andrea}\ \bibnamefont
  {Gambassi}},\ }\bibfield  {title} {\enquote {\bibinfo {title} {Aging and
  coarsening in isolated quantum systems after a quench: Exact results for the
  quantum $\text{O}(n)$ model with $n$ $\ensuremath{\rightarrow}$
  $\ensuremath{\infty}$},}\ }\href {\doibase 10.1103/PhysRevE.92.042151}
  {\bibfield  {journal} {\bibinfo  {journal} {Phys. Rev. E}\ }\textbf {\bibinfo
  {volume} {92}},\ \bibinfo {pages} {042151} (\bibinfo {year}
  {2015})}\BibitemShut {NoStop}%
\bibitem [{\citenamefont {Chiocchetta}\ \emph
  {et~al.}(2016{\natexlab{b}})\citenamefont {Chiocchetta}, \citenamefont
  {Tavora}, \citenamefont {Gambassi},\ and\ \citenamefont
  {Mitra}}]{2016PRB_mitra}%
  \BibitemOpen
  \bibfield  {author} {\bibinfo {author} {\bibfnamefont {Alessio}\ \bibnamefont
  {Chiocchetta}}, \bibinfo {author} {\bibfnamefont {Marco}\ \bibnamefont
  {Tavora}}, \bibinfo {author} {\bibfnamefont {Andrea}\ \bibnamefont
  {Gambassi}}, \ and\ \bibinfo {author} {\bibfnamefont {Aditi}\ \bibnamefont
  {Mitra}},\ }\bibfield  {title} {\enquote {\bibinfo {title} {Short-time
  universal scaling and light-cone dynamics after a quench in an isolated
  quantum system in $d$ spatial dimensions},}\ }\href {\doibase
  10.1103/PhysRevB.94.134311} {\bibfield  {journal} {\bibinfo  {journal} {Phys.
  Rev. B}\ }\textbf {\bibinfo {volume} {94}},\ \bibinfo {pages} {134311}
  (\bibinfo {year} {2016}{\natexlab{b}})}\BibitemShut {NoStop}%
\bibitem [{\citenamefont {{Berges}}(2015)}]{2015bergesreview}%
  \BibitemOpen
  \bibfield  {author} {\bibinfo {author} {\bibfnamefont {J.}~\bibnamefont
  {{Berges}}},\ }\bibfield  {title} {\enquote {\bibinfo {title}
  {{Nonequilibrium Quantum Fields: From Cold Atoms to Cosmology}},}\
  }\href@noop {} {\bibfield  {journal} {\bibinfo  {journal} {arXiv e-prints}\
  ,\ \bibinfo {eid} {arXiv:1503.02907}} (\bibinfo {year} {2015})}\BibitemShut
  {NoStop}%
\bibitem [{\citenamefont {Fujimoto}\ and\ \citenamefont
  {Tsubota}(2016)}]{2016PRA_tsubota}%
  \BibitemOpen
  \bibfield  {author} {\bibinfo {author} {\bibfnamefont {Kazuya}\ \bibnamefont
  {Fujimoto}}\ and\ \bibinfo {author} {\bibfnamefont {Makoto}\ \bibnamefont
  {Tsubota}},\ }\bibfield  {title} {\enquote {\bibinfo {title} {Direct and
  inverse cascades of spin-wave turbulence in spin-1 ferromagnetic spinor
  bose-einstein condensates},}\ }\href {\doibase 10.1103/PhysRevA.93.033620}
  {\bibfield  {journal} {\bibinfo  {journal} {Phys. Rev. A}\ }\textbf {\bibinfo
  {volume} {93}},\ \bibinfo {pages} {033620} (\bibinfo {year}
  {2016})}\BibitemShut {NoStop}%
\bibitem [{\citenamefont {Berges}\ and\ \citenamefont
  {Wallisch}(2017)}]{2017PRD_berges}%
  \BibitemOpen
  \bibfield  {author} {\bibinfo {author} {\bibfnamefont {Jürgen}\ \bibnamefont
  {Berges}}\ and\ \bibinfo {author} {\bibfnamefont {Benjamin}\ \bibnamefont
  {Wallisch}},\ }\bibfield  {title} {\enquote {\bibinfo {title} {{Nonthermal
  Fixed Points in Quantum Field Theory Beyond the Weak-Coupling Limit}},}\
  }\href {\doibase 10.1103/PhysRevD.95.036016} {\bibfield  {journal} {\bibinfo
  {journal} {Phys. Rev.}\ }\textbf {\bibinfo {volume} {D95}},\ \bibinfo {pages}
  {036016} (\bibinfo {year} {2017})}\BibitemShut {NoStop}%
\bibitem [{\citenamefont {Schmied}\ \emph {et~al.}(2019)\citenamefont
  {Schmied}, \citenamefont {Mikheev},\ and\ \citenamefont
  {Gasenzer}}]{2019PRL_gasenzer}%
  \BibitemOpen
  \bibfield  {author} {\bibinfo {author} {\bibfnamefont {Christian-Marcel}\
  \bibnamefont {Schmied}}, \bibinfo {author} {\bibfnamefont {Aleksandr~N.}\
  \bibnamefont {Mikheev}}, \ and\ \bibinfo {author} {\bibfnamefont {Thomas}\
  \bibnamefont {Gasenzer}},\ }\bibfield  {title} {\enquote {\bibinfo {title}
  {Prescaling in a far-from-equilibrium bose gas},}\ }\href {\doibase
  10.1103/PhysRevLett.122.170404} {\bibfield  {journal} {\bibinfo  {journal}
  {Phys. Rev. Lett.}\ }\textbf {\bibinfo {volume} {122}},\ \bibinfo {pages}
  {170404} (\bibinfo {year} {2019})}\BibitemShut {NoStop}%
\bibitem [{\citenamefont {Chiocchetta}\ \emph {et~al.}(2017)\citenamefont
  {Chiocchetta}, \citenamefont {Gambassi}, \citenamefont {Diehl},\ and\
  \citenamefont {Marino}}]{2017PRL_Diehl}%
  \BibitemOpen
  \bibfield  {author} {\bibinfo {author} {\bibfnamefont {Alessio}\ \bibnamefont
  {Chiocchetta}}, \bibinfo {author} {\bibfnamefont {Andrea}\ \bibnamefont
  {Gambassi}}, \bibinfo {author} {\bibfnamefont {Sebastian}\ \bibnamefont
  {Diehl}}, \ and\ \bibinfo {author} {\bibfnamefont {Jamir}\ \bibnamefont
  {Marino}},\ }\bibfield  {title} {\enquote {\bibinfo {title} {Dynamical
  crossovers in prethermal critical states},}\ }\href@noop {} {\bibfield
  {journal} {\bibinfo  {journal} {Physical review letters}\ }\textbf {\bibinfo
  {volume} {118}},\ \bibinfo {pages} {135701} (\bibinfo {year}
  {2017})}\BibitemShut {NoStop}%
\bibitem [{\citenamefont {Navon}\ \emph {et~al.}(2016)\citenamefont {Navon},
  \citenamefont {Gaunt}, \citenamefont {Smith},\ and\ \citenamefont
  {Hadzibabic}}]{2016Nature_hadzibabic}%
  \BibitemOpen
  \bibfield  {author} {\bibinfo {author} {\bibfnamefont {Nir}\ \bibnamefont
  {Navon}}, \bibinfo {author} {\bibfnamefont {Alexander~L.}\ \bibnamefont
  {Gaunt}}, \bibinfo {author} {\bibfnamefont {Robert~P.}\ \bibnamefont
  {Smith}}, \ and\ \bibinfo {author} {\bibfnamefont {Zoran}\ \bibnamefont
  {Hadzibabic}},\ }\bibfield  {title} {\enquote {\bibinfo {title} {Emergence of
  a turbulent cascade in a quantum gas},}\ }\href {\doibase
  10.1038/nature20114} {\bibfield  {journal} {\bibinfo  {journal} {Nature}\
  }\textbf {\bibinfo {volume} {539}},\ \bibinfo {pages} {72--75} (\bibinfo
  {year} {2016})}\BibitemShut {NoStop}%
\bibitem [{\citenamefont {Eigen}\ \emph {et~al.}(2018)\citenamefont {Eigen},
  \citenamefont {Glidden}, \citenamefont {Lopes}, \citenamefont {Cornell},
  \citenamefont {Smith},\ and\ \citenamefont
  {Hadzibabic}}]{2018natureuniversality1}%
  \BibitemOpen
  \bibfield  {author} {\bibinfo {author} {\bibfnamefont {Christoph}\
  \bibnamefont {Eigen}}, \bibinfo {author} {\bibfnamefont {Jake A.~P.}\
  \bibnamefont {Glidden}}, \bibinfo {author} {\bibfnamefont {Raphael}\
  \bibnamefont {Lopes}}, \bibinfo {author} {\bibfnamefont {Eric~A.}\
  \bibnamefont {Cornell}}, \bibinfo {author} {\bibfnamefont {Robert~P.}\
  \bibnamefont {Smith}}, \ and\ \bibinfo {author} {\bibfnamefont {Zoran}\
  \bibnamefont {Hadzibabic}},\ }\bibfield  {title} {\enquote {\bibinfo {title}
  {Universal prethermal dynamics of bose gases quenched to unitarity},}\ }\href
  {https://doi.org/10.1038/s41586-018-0674-1} {\bibfield  {journal} {\bibinfo
  {journal} {Nature}\ }\textbf {\bibinfo {volume} {563}},\ \bibinfo {pages}
  {221--224} (\bibinfo {year} {2018})}\BibitemShut {NoStop}%
\bibitem [{\citenamefont {Erne}\ \emph {et~al.}(2018)\citenamefont {Erne},
  \citenamefont {Bücker}, \citenamefont {Gasenzer}, \citenamefont {Berges},\
  and\ \citenamefont {Schmiedmayer}}]{2018natureuniversality2}%
  \BibitemOpen
  \bibfield  {author} {\bibinfo {author} {\bibfnamefont {Sebastian}\
  \bibnamefont {Erne}}, \bibinfo {author} {\bibfnamefont {Robert}\ \bibnamefont
  {Bücker}}, \bibinfo {author} {\bibfnamefont {Thomas}\ \bibnamefont
  {Gasenzer}}, \bibinfo {author} {\bibfnamefont {Jürgen}\ \bibnamefont
  {Berges}}, \ and\ \bibinfo {author} {\bibfnamefont {Jörg}\ \bibnamefont
  {Schmiedmayer}},\ }\bibfield  {title} {\enquote {\bibinfo {title} {Universal
  dynamics in an isolated one-dimensional bose gas far from equilibrium},}\
  }\href {https://doi.org/10.1038/s41586-018-0667-0} {\bibfield  {journal}
  {\bibinfo  {journal} {Nature}\ }\textbf {\bibinfo {volume} {563}},\ \bibinfo
  {pages} {225--229} (\bibinfo {year} {2018})}\BibitemShut {NoStop}%
\bibitem [{\citenamefont {Prüfer}\ \emph {et~al.}(2018)\citenamefont
  {Prüfer}, \citenamefont {Kunkel}, \citenamefont {Strobel}, \citenamefont
  {Lannig}, \citenamefont {Linnemann}, \citenamefont {Schmied}, \citenamefont
  {Berges}, \citenamefont {Gasenzer},\ and\ \citenamefont
  {Oberthaler}}]{2018natureuniversality3}%
  \BibitemOpen
  \bibfield  {author} {\bibinfo {author} {\bibfnamefont {Maximilian}\
  \bibnamefont {Prüfer}}, \bibinfo {author} {\bibfnamefont {Philipp}\
  \bibnamefont {Kunkel}}, \bibinfo {author} {\bibfnamefont {Helmut}\
  \bibnamefont {Strobel}}, \bibinfo {author} {\bibfnamefont {Stefan}\
  \bibnamefont {Lannig}}, \bibinfo {author} {\bibfnamefont {Daniel}\
  \bibnamefont {Linnemann}}, \bibinfo {author} {\bibfnamefont
  {Christian-Marcel}\ \bibnamefont {Schmied}}, \bibinfo {author} {\bibfnamefont
  {Jürgen}\ \bibnamefont {Berges}}, \bibinfo {author} {\bibfnamefont {Thomas}\
  \bibnamefont {Gasenzer}}, \ and\ \bibinfo {author} {\bibfnamefont
  {Markus~K.}\ \bibnamefont {Oberthaler}},\ }\bibfield  {title} {\enquote
  {\bibinfo {title} {Observation of universal dynamics in a spinor bose gas far
  from equilibrium},}\ }\href {https://doi.org/10.1038/s41586-018-0659-0}
  {\bibfield  {journal} {\bibinfo  {journal} {Nature}\ }\textbf {\bibinfo
  {volume} {563}},\ \bibinfo {pages} {217--220} (\bibinfo {year}
  {2018})}\BibitemShut {NoStop}%
\bibitem [{\citenamefont {{Glidden}}\ \emph {et~al.}(2021)\citenamefont
  {{Glidden}}, \citenamefont {{Eigen}}, \citenamefont {{Dogra}}, \citenamefont
  {{Hilker}}, \citenamefont {{Smith}},\ and\ \citenamefont
  {{Hadzibabic}}}]{2021NP_hadzibabic}%
  \BibitemOpen
  \bibfield  {author} {\bibinfo {author} {\bibfnamefont {Jake A.~P.}\
  \bibnamefont {{Glidden}}}, \bibinfo {author} {\bibfnamefont {Christoph}\
  \bibnamefont {{Eigen}}}, \bibinfo {author} {\bibfnamefont {Lena~H.}\
  \bibnamefont {{Dogra}}}, \bibinfo {author} {\bibfnamefont {Timon~A.}\
  \bibnamefont {{Hilker}}}, \bibinfo {author} {\bibfnamefont {Robert~P.}\
  \bibnamefont {{Smith}}}, \ and\ \bibinfo {author} {\bibfnamefont {Zoran}\
  \bibnamefont {{Hadzibabic}}},\ }\bibfield  {title} {\enquote {\bibinfo
  {title} {{Bidirectional dynamic scaling in an isolated Bose gas far from
  equilibrium}},}\ }\href {\doibase 10.1038/s41567-020-01114-x} {\bibfield
  {journal} {\bibinfo  {journal} {Nature Physics}\ }\textbf {\bibinfo {volume}
  {17}},\ \bibinfo {pages} {457--461} (\bibinfo {year} {2021})}\BibitemShut
  {NoStop}%
\bibitem [{\citenamefont {Bertini}\ \emph {et~al.}(2015)\citenamefont
  {Bertini}, \citenamefont {Essler}, \citenamefont {Groha},\ and\ \citenamefont
  {Robinson}}]{integrabilitybreaking3}%
  \BibitemOpen
  \bibfield  {author} {\bibinfo {author} {\bibfnamefont {Bruno}\ \bibnamefont
  {Bertini}}, \bibinfo {author} {\bibfnamefont {Fabian H.~L.}\ \bibnamefont
  {Essler}}, \bibinfo {author} {\bibfnamefont {Stefan}\ \bibnamefont {Groha}},
  \ and\ \bibinfo {author} {\bibfnamefont {Neil~J.}\ \bibnamefont {Robinson}},\
  }\bibfield  {title} {\enquote {\bibinfo {title} {Prethermalization and
  thermalization in models with weak integrability breaking},}\ }\href
  {\doibase 10.1103/PhysRevLett.115.180601} {\bibfield  {journal} {\bibinfo
  {journal} {Phys. Rev. Lett.}\ }\textbf {\bibinfo {volume} {115}},\ \bibinfo
  {pages} {180601} (\bibinfo {year} {2015})}\BibitemShut {NoStop}%
\bibitem [{\citenamefont {Bulchandani}\ \emph {et~al.}(2018)\citenamefont
  {Bulchandani}, \citenamefont {Vasseur}, \citenamefont {Karrasch},\ and\
  \citenamefont {Moore}}]{2018PRLbulchandani}%
  \BibitemOpen
  \bibfield  {author} {\bibinfo {author} {\bibfnamefont {Vir~B.}\ \bibnamefont
  {Bulchandani}}, \bibinfo {author} {\bibfnamefont {Romain}\ \bibnamefont
  {Vasseur}}, \bibinfo {author} {\bibfnamefont {Christoph}\ \bibnamefont
  {Karrasch}}, \ and\ \bibinfo {author} {\bibfnamefont {Joel~E.}\ \bibnamefont
  {Moore}},\ }\bibfield  {title} {\enquote {\bibinfo {title} {Bethe-{Boltzmann}
  hydrodynamics and spin transport in the {XXZ} chain},}\ }\href {\doibase
  10.1103/PhysRevB.97.045407} {\bibfield  {journal} {\bibinfo  {journal} {Phys.
  Rev. B}\ }\textbf {\bibinfo {volume} {97}},\ \bibinfo {pages} {045407}
  (\bibinfo {year} {2018})}\BibitemShut {NoStop}%
\bibitem [{\citenamefont {De~Nardis}\ \emph {et~al.}(2020)\citenamefont
  {De~Nardis}, \citenamefont {Gopalakrishnan}, \citenamefont {Ilievski},\ and\
  \citenamefont {Vasseur}}]{2020PRL_denardis}%
  \BibitemOpen
  \bibfield  {author} {\bibinfo {author} {\bibfnamefont {Jacopo}\ \bibnamefont
  {De~Nardis}}, \bibinfo {author} {\bibfnamefont {Sarang}\ \bibnamefont
  {Gopalakrishnan}}, \bibinfo {author} {\bibfnamefont {Enej}\ \bibnamefont
  {Ilievski}}, \ and\ \bibinfo {author} {\bibfnamefont {Romain}\ \bibnamefont
  {Vasseur}},\ }\bibfield  {title} {\enquote {\bibinfo {title} {Superdiffusion
  from emergent classical solitons in quantum spin chains},}\ }\href@noop {}
  {\bibfield  {journal} {\bibinfo  {journal} {Physical Review Letters}\
  }\textbf {\bibinfo {volume} {125}},\ \bibinfo {pages} {070601} (\bibinfo
  {year} {2020})}\BibitemShut {NoStop}%
\bibitem [{\citenamefont {Gopalakrishnan}\ and\ \citenamefont
  {Vasseur}(2019)}]{2019PRL_sarang}%
  \BibitemOpen
  \bibfield  {author} {\bibinfo {author} {\bibfnamefont {Sarang}\ \bibnamefont
  {Gopalakrishnan}}\ and\ \bibinfo {author} {\bibfnamefont {Romain}\
  \bibnamefont {Vasseur}},\ }\bibfield  {title} {\enquote {\bibinfo {title}
  {Kinetic theory of spin diffusion and superdiffusion in {$XXZ$} spin
  chains},}\ }\href {\doibase 10.1103/PhysRevLett.122.127202} {\bibfield
  {journal} {\bibinfo  {journal} {Phys. Rev. Lett.}\ }\textbf {\bibinfo
  {volume} {122}},\ \bibinfo {pages} {127202} (\bibinfo {year}
  {2019})}\BibitemShut {NoStop}%
\bibitem [{\citenamefont {Das}\ \emph {et~al.}(2020)\citenamefont {Das},
  \citenamefont {Damle}, \citenamefont {Dhar}, \citenamefont {Huse},
  \citenamefont {Kulkarni}, \citenamefont {Mendl},\ and\ \citenamefont
  {Spohn}}]{das2020nonlinear}%
  \BibitemOpen
  \bibfield  {author} {\bibinfo {author} {\bibfnamefont {Avijit}\ \bibnamefont
  {Das}}, \bibinfo {author} {\bibfnamefont {Kedar}\ \bibnamefont {Damle}},
  \bibinfo {author} {\bibfnamefont {Abhishek}\ \bibnamefont {Dhar}}, \bibinfo
  {author} {\bibfnamefont {David~A}\ \bibnamefont {Huse}}, \bibinfo {author}
  {\bibfnamefont {Manas}\ \bibnamefont {Kulkarni}}, \bibinfo {author}
  {\bibfnamefont {Christian~B}\ \bibnamefont {Mendl}}, \ and\ \bibinfo {author}
  {\bibfnamefont {Herbert}\ \bibnamefont {Spohn}},\ }\bibfield  {title}
  {\enquote {\bibinfo {title} {Nonlinear fluctuating hydrodynamics for the
  classical xxz spin chain},}\ }\href@noop {} {\bibfield  {journal} {\bibinfo
  {journal} {Journal of Statistical Physics}\ }\textbf {\bibinfo {volume}
  {180}},\ \bibinfo {pages} {238--262} (\bibinfo {year} {2020})}\BibitemShut
  {NoStop}%
\bibitem [{\citenamefont {Delacr{\'e}taz}\ and\ \citenamefont
  {Glorioso}(2020)}]{delacretaz2020breakdown}%
  \BibitemOpen
  \bibfield  {author} {\bibinfo {author} {\bibfnamefont {Luca~V}\ \bibnamefont
  {Delacr{\'e}taz}}\ and\ \bibinfo {author} {\bibfnamefont {Paolo}\
  \bibnamefont {Glorioso}},\ }\bibfield  {title} {\enquote {\bibinfo {title}
  {Breakdown of diffusion on chiral edges},}\ }\href@noop {} {\bibfield
  {journal} {\bibinfo  {journal} {Physical Review Letters}\ }\textbf {\bibinfo
  {volume} {124}},\ \bibinfo {pages} {236802} (\bibinfo {year}
  {2020})}\BibitemShut {NoStop}%
\bibitem [{\citenamefont {Glorioso}\ \emph {et~al.}(2022)\citenamefont
  {Glorioso}, \citenamefont {Guo}, \citenamefont {Rodriguez-Nieva},\ and\
  \citenamefont {Lucas}}]{2022NP_glorioso}%
  \BibitemOpen
  \bibfield  {author} {\bibinfo {author} {\bibfnamefont {Paolo}\ \bibnamefont
  {Glorioso}}, \bibinfo {author} {\bibfnamefont {Jinkang}\ \bibnamefont {Guo}},
  \bibinfo {author} {\bibfnamefont {Joaquin~F.}\ \bibnamefont
  {Rodriguez-Nieva}}, \ and\ \bibinfo {author} {\bibfnamefont {Andrew}\
  \bibnamefont {Lucas}},\ }\bibfield  {title} {\enquote {\bibinfo {title}
  {Breakdown of hydrodynamics below four dimensions in a fracton fluid},}\
  }\href {\doibase 10.1038/s41567-022-01631-x} {\bibfield  {journal} {\bibinfo
  {journal} {Nature Physics}\ }\textbf {\bibinfo {volume} {18}},\ \bibinfo
  {pages} {912} (\bibinfo {year} {2022})}\BibitemShut {NoStop}%
\bibitem [{\citenamefont {Rutenberg}(1995)}]{1995rutenberg}%
  \BibitemOpen
  \bibfield  {author} {\bibinfo {author} {\bibfnamefont {A.~D.}\ \bibnamefont
  {Rutenberg}},\ }\bibfield  {title} {\enquote {\bibinfo {title} {Scaling
  violations with textures in two-dimensional phase ordering},}\ }\href
  {\doibase 10.1103/PhysRevE.51.R2715} {\bibfield  {journal} {\bibinfo
  {journal} {Phys. Rev. E}\ }\textbf {\bibinfo {volume} {51}},\ \bibinfo
  {pages} {R2715--R2718} (\bibinfo {year} {1995})}\BibitemShut {NoStop}%
\bibitem [{\citenamefont {Rutenberg}\ and\ \citenamefont
  {Bray}(1995)}]{1995PRE_bray}%
  \BibitemOpen
  \bibfield  {author} {\bibinfo {author} {\bibfnamefont {A.~D.}\ \bibnamefont
  {Rutenberg}}\ and\ \bibinfo {author} {\bibfnamefont {A.~J.}\ \bibnamefont
  {Bray}},\ }\bibfield  {title} {\enquote {\bibinfo {title} {Energy-scaling
  approach to phase-ordering growth laws},}\ }\href {\doibase
  10.1103/PhysRevE.51.5499} {\bibfield  {journal} {\bibinfo  {journal} {Phys.
  Rev. E}\ }\textbf {\bibinfo {volume} {51}},\ \bibinfo {pages} {5499--5514}
  (\bibinfo {year} {1995})}\BibitemShut {NoStop}%
\bibitem [{\citenamefont {Gagel}\ \emph {et~al.}(2014)\citenamefont {Gagel},
  \citenamefont {Orth},\ and\ \citenamefont {Schmalian}}]{2014PRL_schmalian}%
  \BibitemOpen
  \bibfield  {author} {\bibinfo {author} {\bibfnamefont {Pia}\ \bibnamefont
  {Gagel}}, \bibinfo {author} {\bibfnamefont {Peter~P}\ \bibnamefont {Orth}}, \
  and\ \bibinfo {author} {\bibfnamefont {J{\"o}rg}\ \bibnamefont {Schmalian}},\
  }\bibfield  {title} {\enquote {\bibinfo {title} {Universal postquench
  prethermalization at a quantum critical point},}\ }\href@noop {} {\bibfield
  {journal} {\bibinfo  {journal} {Physical review letters}\ }\textbf {\bibinfo
  {volume} {113}},\ \bibinfo {pages} {220401} (\bibinfo {year}
  {2014})}\BibitemShut {NoStop}%
\bibitem [{\citenamefont {Gagel}\ \emph {et~al.}(2015)\citenamefont {Gagel},
  \citenamefont {Orth},\ and\ \citenamefont {Schmalian}}]{2015PRB_schmalian}%
  \BibitemOpen
  \bibfield  {author} {\bibinfo {author} {\bibfnamefont {Pia}\ \bibnamefont
  {Gagel}}, \bibinfo {author} {\bibfnamefont {Peter~P}\ \bibnamefont {Orth}}, \
  and\ \bibinfo {author} {\bibfnamefont {J{\"o}rg}\ \bibnamefont {Schmalian}},\
  }\bibfield  {title} {\enquote {\bibinfo {title} {Universal postquench
  coarsening and aging at a quantum critical point},}\ }\href@noop {}
  {\bibfield  {journal} {\bibinfo  {journal} {Physical Review B}\ }\textbf
  {\bibinfo {volume} {92}},\ \bibinfo {pages} {115121} (\bibinfo {year}
  {2015})}\BibitemShut {NoStop}%
\bibitem [{\citenamefont {Sciolla}\ and\ \citenamefont
  {Biroli}(2013)}]{2013PRB_dynamicaltransitions}%
  \BibitemOpen
  \bibfield  {author} {\bibinfo {author} {\bibfnamefont {Bruno}\ \bibnamefont
  {Sciolla}}\ and\ \bibinfo {author} {\bibfnamefont {Giulio}\ \bibnamefont
  {Biroli}},\ }\bibfield  {title} {\enquote {\bibinfo {title} {Quantum
  quenches, dynamical transitions, and off-equilibrium quantum criticality},}\
  }\href@noop {} {\bibfield  {journal} {\bibinfo  {journal} {Physical Review
  B}\ }\textbf {\bibinfo {volume} {88}},\ \bibinfo {pages} {201110} (\bibinfo
  {year} {2013})}\BibitemShut {NoStop}%
\bibitem [{\citenamefont {Boguslavski}\ and\ \citenamefont {Pi\~neiro
  Orioli}(2020)}]{2020_Onasier}%
  \BibitemOpen
  \bibfield  {author} {\bibinfo {author} {\bibfnamefont {Kirill}\ \bibnamefont
  {Boguslavski}}\ and\ \bibinfo {author} {\bibfnamefont {Asier}\ \bibnamefont
  {Pi\~neiro Orioli}},\ }\bibfield  {title} {\enquote {\bibinfo {title}
  {Unraveling the nature of universal dynamics in {$O(N)$} theories},}\ }\href
  {\doibase 10.1103/PhysRevD.101.091902} {\bibfield  {journal} {\bibinfo
  {journal} {Phys. Rev. D}\ }\textbf {\bibinfo {volume} {101}},\ \bibinfo
  {pages} {091902} (\bibinfo {year} {2020})}\BibitemShut {NoStop}%
\bibitem [{\citenamefont {Bhattacharyya}\ \emph {et~al.}(2020)\citenamefont
  {Bhattacharyya}, \citenamefont {Rodriguez-Nieva},\ and\ \citenamefont
  {Demler}}]{2020PRL_rodrigueznieva}%
  \BibitemOpen
  \bibfield  {author} {\bibinfo {author} {\bibfnamefont {Saraswat}\
  \bibnamefont {Bhattacharyya}}, \bibinfo {author} {\bibfnamefont {Joaquin~F.}\
  \bibnamefont {Rodriguez-Nieva}}, \ and\ \bibinfo {author} {\bibfnamefont
  {Eugene}\ \bibnamefont {Demler}},\ }\bibfield  {title} {\enquote {\bibinfo
  {title} {Universal prethermal dynamics in heisenberg ferromagnets},}\ }\href
  {\doibase 10.1103/PhysRevLett.125.230601} {\bibfield  {journal} {\bibinfo
  {journal} {Phys. Rev. Lett.}\ }\textbf {\bibinfo {volume} {125}},\ \bibinfo
  {pages} {230601} (\bibinfo {year} {2020})}\BibitemShut {NoStop}%
\bibitem [{\citenamefont {Rodriguez-Nieva}\ \emph
  {et~al.}(2022{\natexlab{a}})\citenamefont {Rodriguez-Nieva}, \citenamefont
  {Orioli},\ and\ \citenamefont {Marino}}]{2022PNAS_rodrigueznieva}%
  \BibitemOpen
  \bibfield  {author} {\bibinfo {author} {\bibfnamefont {Joaquin~F.}\
  \bibnamefont {Rodriguez-Nieva}}, \bibinfo {author} {\bibfnamefont
  {Asier~Piñeiro}\ \bibnamefont {Orioli}}, \ and\ \bibinfo {author}
  {\bibfnamefont {Jamir}\ \bibnamefont {Marino}},\ }\bibfield  {title}
  {\enquote {\bibinfo {title} {Far-from-equilibrium universality in the
  two-dimensional heisenberg model},}\ }\href {\doibase
  10.1073/pnas.2122599119} {\bibfield  {journal} {\bibinfo  {journal}
  {Proceedings of the National Academy of Sciences}\ }\textbf {\bibinfo
  {volume} {119}},\ \bibinfo {pages} {e2122599119} (\bibinfo {year}
  {2022}{\natexlab{a}})}\BibitemShut {NoStop}%
\bibitem [{\citenamefont {Hohenberg}\ and\ \citenamefont
  {Halperin}(1977)}]{1977PR_HH}%
  \BibitemOpen
  \bibfield  {author} {\bibinfo {author} {\bibfnamefont {P.~C.}\ \bibnamefont
  {Hohenberg}}\ and\ \bibinfo {author} {\bibfnamefont {B.~I.}\ \bibnamefont
  {Halperin}},\ }\bibfield  {title} {\enquote {\bibinfo {title} {Theory of
  dynamic critical phenomena},}\ }\href {\doibase 10.1103/RevModPhys.49.435}
  {\bibfield  {journal} {\bibinfo  {journal} {Rev. Mod. Phys.}\ }\textbf
  {\bibinfo {volume} {49}},\ \bibinfo {pages} {435--479} (\bibinfo {year}
  {1977})}\BibitemShut {NoStop}%
\bibitem [{\citenamefont {Sachdev}(2011)}]{booksachdev}%
  \BibitemOpen
  \bibfield  {author} {\bibinfo {author} {\bibfnamefont {Subir}\ \bibnamefont
  {Sachdev}},\ }\href {\doibase 10.1017/CBO9780511973765} {\emph {\bibinfo
  {title} {Quantum Phase Transitions}}},\ \bibinfo {edition} {2nd}\ ed.\
  (\bibinfo  {publisher} {Cambridge University Press},\ \bibinfo {year}
  {2011})\BibitemShut {NoStop}%
\bibitem [{\citenamefont {Pi\~neiro Orioli}\ \emph {et~al.}(2015)\citenamefont
  {Pi\~neiro Orioli}, \citenamefont {Boguslavski},\ and\ \citenamefont
  {Berges}}]{2015asier}%
  \BibitemOpen
  \bibfield  {author} {\bibinfo {author} {\bibfnamefont {Asier}\ \bibnamefont
  {Pi\~neiro Orioli}}, \bibinfo {author} {\bibfnamefont {Kirill}\ \bibnamefont
  {Boguslavski}}, \ and\ \bibinfo {author} {\bibfnamefont {J\"urgen}\
  \bibnamefont {Berges}},\ }\bibfield  {title} {\enquote {\bibinfo {title}
  {Universal self-similar dynamics of relativistic and nonrelativistic field
  theories near nonthermal fixed points},}\ }\href {\doibase
  10.1103/PhysRevD.92.025041} {\bibfield  {journal} {\bibinfo  {journal} {Phys.
  Rev. D}\ }\textbf {\bibinfo {volume} {92}},\ \bibinfo {pages} {025041}
  (\bibinfo {year} {2015})}\BibitemShut {NoStop}%
\bibitem [{\citenamefont {Hild}\ \emph {et~al.}(2014)\citenamefont {Hild},
  \citenamefont {Fukuhara}, \citenamefont {Schau\ss{}}, \citenamefont {Zeiher},
  \citenamefont {Knap}, \citenamefont {Demler}, \citenamefont {Bloch},\ and\
  \citenamefont {Gross}}]{2014spiralstate}%
  \BibitemOpen
  \bibfield  {author} {\bibinfo {author} {\bibfnamefont {Sebastian}\
  \bibnamefont {Hild}}, \bibinfo {author} {\bibfnamefont {Takeshi}\
  \bibnamefont {Fukuhara}}, \bibinfo {author} {\bibfnamefont {Peter}\
  \bibnamefont {Schau\ss{}}}, \bibinfo {author} {\bibfnamefont {Johannes}\
  \bibnamefont {Zeiher}}, \bibinfo {author} {\bibfnamefont {Michael}\
  \bibnamefont {Knap}}, \bibinfo {author} {\bibfnamefont {Eugene}\ \bibnamefont
  {Demler}}, \bibinfo {author} {\bibfnamefont {Immanuel}\ \bibnamefont
  {Bloch}}, \ and\ \bibinfo {author} {\bibfnamefont {Christian}\ \bibnamefont
  {Gross}},\ }\bibfield  {title} {\enquote {\bibinfo {title}
  {Far-from-equilibrium spin transport in heisenberg quantum magnets},}\ }\href
  {\doibase 10.1103/PhysRevLett.113.147205} {\bibfield  {journal} {\bibinfo
  {journal} {Phys. Rev. Lett.}\ }\textbf {\bibinfo {volume} {113}},\ \bibinfo
  {pages} {147205} (\bibinfo {year} {2014})}\BibitemShut {NoStop}%
\bibitem [{\citenamefont {Jepsen}\ \emph {et~al.}(2020)\citenamefont {Jepsen},
  \citenamefont {Amato-Grill}, \citenamefont {Dimitrova}, \citenamefont {Ho},
  \citenamefont {Demler},\ and\ \citenamefont {Ketterle}}]{2020Nature_jepsen}%
  \BibitemOpen
  \bibfield  {author} {\bibinfo {author} {\bibfnamefont {Paul~Niklas}\
  \bibnamefont {Jepsen}}, \bibinfo {author} {\bibfnamefont {Jesse}\
  \bibnamefont {Amato-Grill}}, \bibinfo {author} {\bibfnamefont {Ivana}\
  \bibnamefont {Dimitrova}}, \bibinfo {author} {\bibfnamefont {Wen~Wei}\
  \bibnamefont {Ho}}, \bibinfo {author} {\bibfnamefont {Eugene}\ \bibnamefont
  {Demler}}, \ and\ \bibinfo {author} {\bibfnamefont {Wolfgang}\ \bibnamefont
  {Ketterle}},\ }\bibfield  {title} {\enquote {\bibinfo {title} {Spin transport
  in a tunable heisenberg model realized with ultracold atoms},}\ }\href
  {\doibase 10.1038/s41586-020-3033-y} {\bibfield  {journal} {\bibinfo
  {journal} {Nature}\ }\textbf {\bibinfo {volume} {588}},\ \bibinfo {pages}
  {403--407} (\bibinfo {year} {2020})}\BibitemShut {NoStop}%
\bibitem [{\citenamefont {Jepsen}\ \emph {et~al.}(2021)\citenamefont {Jepsen},
  \citenamefont {Ho}, \citenamefont {Amato-Grill}, \citenamefont {Dimitrova},
  \citenamefont {Demler},\ and\ \citenamefont {Ketterle}}]{2021PRX_jepsen}%
  \BibitemOpen
  \bibfield  {author} {\bibinfo {author} {\bibfnamefont {Paul~Niklas}\
  \bibnamefont {Jepsen}}, \bibinfo {author} {\bibfnamefont {Wen~Wei}\
  \bibnamefont {Ho}}, \bibinfo {author} {\bibfnamefont {Jesse}\ \bibnamefont
  {Amato-Grill}}, \bibinfo {author} {\bibfnamefont {Ivana}\ \bibnamefont
  {Dimitrova}}, \bibinfo {author} {\bibfnamefont {Eugene}\ \bibnamefont
  {Demler}}, \ and\ \bibinfo {author} {\bibfnamefont {Wolfgang}\ \bibnamefont
  {Ketterle}},\ }\bibfield  {title} {\enquote {\bibinfo {title} {Transverse
  spin dynamics in the anisotropic heisenberg model realized with ultracold
  atoms},}\ }\href@noop {} {\bibfield  {journal} {\bibinfo  {journal} {arXiv
  preprint arXiv:2103.07866}\ } (\bibinfo {year} {2021})}\BibitemShut {NoStop}%
\bibitem [{\citenamefont {Du}\ \emph {et~al.}(2017)\citenamefont {Du},
  \citenamefont {van~der Sar}, \citenamefont {Zhou}, \citenamefont {Upadhyaya},
  \citenamefont {Casola}, \citenamefont {Zhang}, \citenamefont {Onbasli},
  \citenamefont {Ross}, \citenamefont {Walsworth}, \citenamefont
  {Tserkovnyak},\ and\ \citenamefont {Yacoby}}]{2017chunhui}%
  \BibitemOpen
  \bibfield  {author} {\bibinfo {author} {\bibfnamefont {Chunhui}\ \bibnamefont
  {Du}}, \bibinfo {author} {\bibfnamefont {Toeno}\ \bibnamefont {van~der Sar}},
  \bibinfo {author} {\bibfnamefont {Tony~X.}\ \bibnamefont {Zhou}}, \bibinfo
  {author} {\bibfnamefont {Pramey}\ \bibnamefont {Upadhyaya}}, \bibinfo
  {author} {\bibfnamefont {Francesco}\ \bibnamefont {Casola}}, \bibinfo
  {author} {\bibfnamefont {Huiliang}\ \bibnamefont {Zhang}}, \bibinfo {author}
  {\bibfnamefont {Mehmet~C.}\ \bibnamefont {Onbasli}}, \bibinfo {author}
  {\bibfnamefont {Caroline~A.}\ \bibnamefont {Ross}}, \bibinfo {author}
  {\bibfnamefont {Ronald~L.}\ \bibnamefont {Walsworth}}, \bibinfo {author}
  {\bibfnamefont {Yaroslav}\ \bibnamefont {Tserkovnyak}}, \ and\ \bibinfo
  {author} {\bibfnamefont {Amir}\ \bibnamefont {Yacoby}},\ }\bibfield  {title}
  {\enquote {\bibinfo {title} {Control and local measurement of the spin
  chemical potential in a magnetic insulator},}\ }\href {\doibase
  10.1126/science.aak9611} {\bibfield  {journal} {\bibinfo  {journal}
  {Science}\ }\textbf {\bibinfo {volume} {357}},\ \bibinfo {pages} {195--198}
  (\bibinfo {year} {2017})}\BibitemShut {NoStop}%
\bibitem [{\citenamefont {Rodriguez-Nieva}\ \emph
  {et~al.}(2022{\natexlab{b}})\citenamefont {Rodriguez-Nieva}, \citenamefont
  {Podolsky},\ and\ \citenamefont {Demler}}]{2018nv-ferro}%
  \BibitemOpen
  \bibfield  {author} {\bibinfo {author} {\bibfnamefont {Joaquin~F.}\
  \bibnamefont {Rodriguez-Nieva}}, \bibinfo {author} {\bibfnamefont {Daniel}\
  \bibnamefont {Podolsky}}, \ and\ \bibinfo {author} {\bibfnamefont {Eugene}\
  \bibnamefont {Demler}},\ }\bibfield  {title} {\enquote {\bibinfo {title}
  {Probing hydrodynamic sound modes in magnon fluids using spin
  magnetometers},}\ }\href {\doibase 10.1103/PhysRevB.105.174412} {\bibfield
  {journal} {\bibinfo  {journal} {Phys. Rev. B}\ }\textbf {\bibinfo {volume}
  {105}},\ \bibinfo {pages} {174412} (\bibinfo {year}
  {2022}{\natexlab{b}})}\BibitemShut {NoStop}%
\bibitem [{\citenamefont {Zhou}\ \emph {et~al.}(2021)\citenamefont {Zhou},
  \citenamefont {Carmiggelt}, \citenamefont {Gächter}, \citenamefont
  {Esterlis}, \citenamefont {Sels}, \citenamefont {Stöhr}, \citenamefont {Du},
  \citenamefont {Fernandez}, \citenamefont {Rodriguez-Nieva}, \citenamefont
  {Büttner}, \citenamefont {Demler},\ and\ \citenamefont
  {Yacoby}}]{2021pnas_tony}%
  \BibitemOpen
  \bibfield  {author} {\bibinfo {author} {\bibfnamefont {Tony~X.}\ \bibnamefont
  {Zhou}}, \bibinfo {author} {\bibfnamefont {Joris~J.}\ \bibnamefont
  {Carmiggelt}}, \bibinfo {author} {\bibfnamefont {Lisa~M.}\ \bibnamefont
  {Gächter}}, \bibinfo {author} {\bibfnamefont {Ilya}\ \bibnamefont
  {Esterlis}}, \bibinfo {author} {\bibfnamefont {Dries}\ \bibnamefont {Sels}},
  \bibinfo {author} {\bibfnamefont {Rainer~J.}\ \bibnamefont {Stöhr}},
  \bibinfo {author} {\bibfnamefont {Chunhui}\ \bibnamefont {Du}}, \bibinfo
  {author} {\bibfnamefont {Daniel}\ \bibnamefont {Fernandez}}, \bibinfo
  {author} {\bibfnamefont {Joaquin~F.}\ \bibnamefont {Rodriguez-Nieva}},
  \bibinfo {author} {\bibfnamefont {Felix}\ \bibnamefont {Büttner}}, \bibinfo
  {author} {\bibfnamefont {Eugene}\ \bibnamefont {Demler}}, \ and\ \bibinfo
  {author} {\bibfnamefont {Amir}\ \bibnamefont {Yacoby}},\ }\bibfield  {title}
  {\enquote {\bibinfo {title} {A magnon scattering platform},}\ }\href
  {\doibase 10.1073/pnas.2019473118} {\bibfield  {journal} {\bibinfo  {journal}
  {Proceedings of the National Academy of Sciences}\ }\textbf {\bibinfo
  {volume} {118}},\ \bibinfo {pages} {e2019473118} (\bibinfo {year}
  {2021})}\BibitemShut {NoStop}%
\bibitem [{\citenamefont {Lee-Wong}\ \emph {et~al.}(2021)\citenamefont
  {Lee-Wong}, \citenamefont {Ding}, \citenamefont {Wang}, \citenamefont {Liu},
  \citenamefont {McLaughlin}, \citenamefont {Wang}, \citenamefont {Wu},\ and\
  \citenamefont {Du}}]{2021PRApplied_chunhui}%
  \BibitemOpen
  \bibfield  {author} {\bibinfo {author} {\bibfnamefont {Eric}\ \bibnamefont
  {Lee-Wong}}, \bibinfo {author} {\bibfnamefont {Jinjun}\ \bibnamefont {Ding}},
  \bibinfo {author} {\bibfnamefont {Xiaoche}\ \bibnamefont {Wang}}, \bibinfo
  {author} {\bibfnamefont {Chuanpu}\ \bibnamefont {Liu}}, \bibinfo {author}
  {\bibfnamefont {Nathan~J.}\ \bibnamefont {McLaughlin}}, \bibinfo {author}
  {\bibfnamefont {Hailong}\ \bibnamefont {Wang}}, \bibinfo {author}
  {\bibfnamefont {Mingzhong}\ \bibnamefont {Wu}}, \ and\ \bibinfo {author}
  {\bibfnamefont {Chunhui~Rita}\ \bibnamefont {Du}},\ }\bibfield  {title}
  {\enquote {\bibinfo {title} {Quantum sensing of spin fluctuations of magnetic
  insulator films with perpendicular anisotropy},}\ }\href {\doibase
  10.1103/PhysRevApplied.15.034031} {\bibfield  {journal} {\bibinfo  {journal}
  {Phys. Rev. Applied}\ }\textbf {\bibinfo {volume} {15}},\ \bibinfo {pages}
  {034031} (\bibinfo {year} {2021})}\BibitemShut {NoStop}%
\bibitem [{\citenamefont {Wang}\ \emph {et~al.}(2021)\citenamefont {Wang},
  \citenamefont {Xiao}, \citenamefont {Guo}, \citenamefont {Lee-Wong},
  \citenamefont {Yan}, \citenamefont {Cheng},\ and\ \citenamefont
  {Du}}]{2021PRL_chunhui}%
  \BibitemOpen
  \bibfield  {author} {\bibinfo {author} {\bibfnamefont {Hailong}\ \bibnamefont
  {Wang}}, \bibinfo {author} {\bibfnamefont {Yuxuan}\ \bibnamefont {Xiao}},
  \bibinfo {author} {\bibfnamefont {Mingda}\ \bibnamefont {Guo}}, \bibinfo
  {author} {\bibfnamefont {Eric}\ \bibnamefont {Lee-Wong}}, \bibinfo {author}
  {\bibfnamefont {Gerald~Q.}\ \bibnamefont {Yan}}, \bibinfo {author}
  {\bibfnamefont {Ran}\ \bibnamefont {Cheng}}, \ and\ \bibinfo {author}
  {\bibfnamefont {Chunhui~Rita}\ \bibnamefont {Du}},\ }\bibfield  {title}
  {\enquote {\bibinfo {title} {Spin pumping of an easy-plane antiferromagnet
  enhanced by dzyaloshinskii--moriya interaction},}\ }\href {\doibase
  10.1103/PhysRevLett.127.117202} {\bibfield  {journal} {\bibinfo  {journal}
  {Phys. Rev. Lett.}\ }\textbf {\bibinfo {volume} {127}},\ \bibinfo {pages}
  {117202} (\bibinfo {year} {2021})}\BibitemShut {NoStop}%
\bibitem [{\citenamefont {Huse}\ and\ \citenamefont
  {Elser}(1988)}]{1988PRL_afmgroundstate}%
  \BibitemOpen
  \bibfield  {author} {\bibinfo {author} {\bibfnamefont {David~A.}\
  \bibnamefont {Huse}}\ and\ \bibinfo {author} {\bibfnamefont {Veit}\
  \bibnamefont {Elser}},\ }\bibfield  {title} {\enquote {\bibinfo {title}
  {Simple variational wave functions for two-dimensional heisenberg
  spin-1/2 antiferromagnets},}\ }\href {\doibase
  10.1103/PhysRevLett.60.2531} {\bibfield  {journal} {\bibinfo  {journal}
  {Phys. Rev. Lett.}\ }\textbf {\bibinfo {volume} {60}},\ \bibinfo {pages}
  {2531--2534} (\bibinfo {year} {1988})}\BibitemShut {NoStop}%
\bibitem [{\citenamefont {Reger}\ and\ \citenamefont
  {Young}(1988)}]{1988PRB_afmgroundstate}%
  \BibitemOpen
  \bibfield  {author} {\bibinfo {author} {\bibfnamefont {J.~D.}\ \bibnamefont
  {Reger}}\ and\ \bibinfo {author} {\bibfnamefont {A.~P.}\ \bibnamefont
  {Young}},\ }\bibfield  {title} {\enquote {\bibinfo {title} {Monte carlo
  simulations of the spin-(1/2 heisenberg antiferromagnet on a square
  lattice},}\ }\href {\doibase 10.1103/PhysRevB.37.5978} {\bibfield  {journal}
  {\bibinfo  {journal} {Phys. Rev. B}\ }\textbf {\bibinfo {volume} {37}},\
  \bibinfo {pages} {5978--5981} (\bibinfo {year} {1988})}\BibitemShut {NoStop}%
\bibitem [{\citenamefont {Babadi}\ \emph {et~al.}(2015)\citenamefont {Babadi},
  \citenamefont {Demler},\ and\ \citenamefont {Knap}}]{2015PRX-babadi}%
  \BibitemOpen
  \bibfield  {author} {\bibinfo {author} {\bibfnamefont {Mehrtash}\
  \bibnamefont {Babadi}}, \bibinfo {author} {\bibfnamefont {Eugene}\
  \bibnamefont {Demler}}, \ and\ \bibinfo {author} {\bibfnamefont {Michael}\
  \bibnamefont {Knap}},\ }\bibfield  {title} {\enquote {\bibinfo {title}
  {Far-from-equilibrium field theory of many-body quantum spin systems:
  Prethermalization and relaxation of spin spiral states in three
  dimensions},}\ }\href {\doibase 10.1103/PhysRevX.5.041005} {\bibfield
  {journal} {\bibinfo  {journal} {Phys. Rev. X}\ }\textbf {\bibinfo {volume}
  {5}},\ \bibinfo {pages} {041005} (\bibinfo {year} {2015})}\BibitemShut
  {NoStop}%
\bibitem [{\citenamefont {Rodriguez-Nieva}\ \emph
  {et~al.}(2022{\natexlab{c}})\citenamefont {Rodriguez-Nieva}, \citenamefont
  {Schuckert}, \citenamefont {Sels}, \citenamefont {Knap},\ and\ \citenamefont
  {Demler}}]{2020PRBspinspiral}%
  \BibitemOpen
  \bibfield  {author} {\bibinfo {author} {\bibfnamefont {Joaquin~F.}\
  \bibnamefont {Rodriguez-Nieva}}, \bibinfo {author} {\bibfnamefont
  {Alexander}\ \bibnamefont {Schuckert}}, \bibinfo {author} {\bibfnamefont
  {Dries}\ \bibnamefont {Sels}}, \bibinfo {author} {\bibfnamefont {Michael}\
  \bibnamefont {Knap}}, \ and\ \bibinfo {author} {\bibfnamefont {Eugene}\
  \bibnamefont {Demler}},\ }\bibfield  {title} {\enquote {\bibinfo {title}
  {Transverse instability and universal decay of spin spiral order in the
  heisenberg model},}\ }\href {\doibase 10.1103/PhysRevB.105.L060302}
  {\bibfield  {journal} {\bibinfo  {journal} {Phys. Rev. B}\ }\textbf {\bibinfo
  {volume} {105}},\ \bibinfo {pages} {L060302} (\bibinfo {year}
  {2022}{\natexlab{c}})}\BibitemShut {NoStop}%
\bibitem [{\citenamefont {Rodriguez-Nieva}(2021)}]{2020PRB_spinturbulence}%
  \BibitemOpen
  \bibfield  {author} {\bibinfo {author} {\bibfnamefont {Joaquin~F.}\
  \bibnamefont {Rodriguez-Nieva}},\ }\bibfield  {title} {\enquote {\bibinfo
  {title} {Turbulent relaxation after a quench in the heisenberg model},}\
  }\href {\doibase 10.1103/PhysRevB.104.L060302} {\bibfield  {journal}
  {\bibinfo  {journal} {Phys. Rev. B}\ }\textbf {\bibinfo {volume} {104}},\
  \bibinfo {pages} {L060302} (\bibinfo {year} {2021})}\BibitemShut {NoStop}%
\bibitem [{\citenamefont {Harris}\ \emph {et~al.}(1971)\citenamefont {Harris},
  \citenamefont {Kumar}, \citenamefont {Halperin},\ and\ \citenamefont
  {Hohenberg}}]{1971PRB_AFMhh}%
  \BibitemOpen
  \bibfield  {author} {\bibinfo {author} {\bibfnamefont {A.~B.}\ \bibnamefont
  {Harris}}, \bibinfo {author} {\bibfnamefont {D.}~\bibnamefont {Kumar}},
  \bibinfo {author} {\bibfnamefont {B.~I.}\ \bibnamefont {Halperin}}, \ and\
  \bibinfo {author} {\bibfnamefont {P.~C.}\ \bibnamefont {Hohenberg}},\
  }\bibfield  {title} {\enquote {\bibinfo {title} {Dynamics of an
  antiferromagnet at low temperatures: Spin-wave damping and hydrodynamics},}\
  }\href {\doibase 10.1103/PhysRevB.3.961} {\bibfield  {journal} {\bibinfo
  {journal} {Phys. Rev. B}\ }\textbf {\bibinfo {volume} {3}},\ \bibinfo {pages}
  {961--1024} (\bibinfo {year} {1971})}\BibitemShut {NoStop}%
\bibitem [{\citenamefont {Holstein}\ and\ \citenamefont
  {Primakoff}(1940)}]{1940PR_holteinprimakoff}%
  \BibitemOpen
  \bibfield  {author} {\bibinfo {author} {\bibfnamefont {T.}~\bibnamefont
  {Holstein}}\ and\ \bibinfo {author} {\bibfnamefont {H.}~\bibnamefont
  {Primakoff}},\ }\bibfield  {title} {\enquote {\bibinfo {title} {Field
  dependence of the intrinsic domain magnetization of a ferromagnet},}\ }\href
  {\doibase 10.1103/PhysRev.58.1098} {\bibfield  {journal} {\bibinfo  {journal}
  {Phys. Rev.}\ }\textbf {\bibinfo {volume} {58}},\ \bibinfo {pages}
  {1098--1113} (\bibinfo {year} {1940})}\BibitemShut {NoStop}%
\bibitem [{\citenamefont {Canali}\ \emph {et~al.}(1992)\citenamefont {Canali},
  \citenamefont {Girvin},\ and\ \citenamefont
  {Wallin}}]{1992PRB_afmdysonmaleev}%
  \BibitemOpen
  \bibfield  {author} {\bibinfo {author} {\bibfnamefont {C.~M.}\ \bibnamefont
  {Canali}}, \bibinfo {author} {\bibfnamefont {S.~M.}\ \bibnamefont {Girvin}},
  \ and\ \bibinfo {author} {\bibfnamefont {Mats}\ \bibnamefont {Wallin}},\
  }\bibfield  {title} {\enquote {\bibinfo {title} {Spin-wave velocity
  renormalization in the two-dimensional heisenberg antiferromagnet at zero
  temperature},}\ }\href {\doibase 10.1103/PhysRevB.45.10131} {\bibfield
  {journal} {\bibinfo  {journal} {Phys. Rev. B}\ }\textbf {\bibinfo {volume}
  {45}},\ \bibinfo {pages} {10131--10134} (\bibinfo {year} {1992})}\BibitemShut
  {NoStop}%
\bibitem [{\citenamefont {Hamer}\ \emph {et~al.}(1992)\citenamefont {Hamer},
  \citenamefont {Weihong},\ and\ \citenamefont
  {Arndt}}]{1992PRB_afmdysonmaleev1}%
  \BibitemOpen
  \bibfield  {author} {\bibinfo {author} {\bibfnamefont {C.~J.}\ \bibnamefont
  {Hamer}}, \bibinfo {author} {\bibfnamefont {Zheng}\ \bibnamefont {Weihong}},
  \ and\ \bibinfo {author} {\bibfnamefont {Peter}\ \bibnamefont {Arndt}},\
  }\bibfield  {title} {\enquote {\bibinfo {title} {Third-order spin-wave theory
  for the heisenberg antiferromagnet},}\ }\href {\doibase
  10.1103/PhysRevB.46.6276} {\bibfield  {journal} {\bibinfo  {journal} {Phys.
  Rev. B}\ }\textbf {\bibinfo {volume} {46}},\ \bibinfo {pages} {6276--6292}
  (\bibinfo {year} {1992})}\BibitemShut {NoStop}%
\bibitem [{\citenamefont {Polkovnikov}(2010)}]{2010review_polkovnikov}%
  \BibitemOpen
  \bibfield  {author} {\bibinfo {author} {\bibfnamefont {Anatoli}\ \bibnamefont
  {Polkovnikov}},\ }\bibfield  {title} {\enquote {\bibinfo {title} {Phase space
  representation of quantum dynamics},}\ }\href@noop {} {\bibfield  {journal}
  {\bibinfo  {journal} {Annals of Physics}\ }\textbf {\bibinfo {volume}
  {325}},\ \bibinfo {pages} {1790--1852} (\bibinfo {year} {2010})}\BibitemShut
  {NoStop}%
\bibitem [{\citenamefont {Davidson}\ and\ \citenamefont
  {Polkovnikov}(2015)}]{spintwa1}%
  \BibitemOpen
  \bibfield  {author} {\bibinfo {author} {\bibfnamefont {Shainen~M.}\
  \bibnamefont {Davidson}}\ and\ \bibinfo {author} {\bibfnamefont {Anatoli}\
  \bibnamefont {Polkovnikov}},\ }\bibfield  {title} {\enquote {\bibinfo {title}
  {$su(3)$ semiclassical representation of quantum dynamics of interacting
  spins},}\ }\href {\doibase 10.1103/PhysRevLett.114.045701} {\bibfield
  {journal} {\bibinfo  {journal} {Phys. Rev. Lett.}\ }\textbf {\bibinfo
  {volume} {114}},\ \bibinfo {pages} {045701} (\bibinfo {year}
  {2015})}\BibitemShut {NoStop}%
\bibitem [{\citenamefont {Schachenmayer}\ \emph {et~al.}(2015)\citenamefont
  {Schachenmayer}, \citenamefont {Pikovski},\ and\ \citenamefont
  {Rey}}]{spintwa2}%
  \BibitemOpen
  \bibfield  {author} {\bibinfo {author} {\bibfnamefont {J.}~\bibnamefont
  {Schachenmayer}}, \bibinfo {author} {\bibfnamefont {A.}~\bibnamefont
  {Pikovski}}, \ and\ \bibinfo {author} {\bibfnamefont {A.~M.}\ \bibnamefont
  {Rey}},\ }\bibfield  {title} {\enquote {\bibinfo {title} {Many-body quantum
  spin dynamics with monte carlo trajectories on a discrete phase space},}\
  }\href {\doibase 10.1103/PhysRevX.5.011022} {\bibfield  {journal} {\bibinfo
  {journal} {Phys. Rev. X}\ }\textbf {\bibinfo {volume} {5}},\ \bibinfo {pages}
  {011022} (\bibinfo {year} {2015})}\BibitemShut {NoStop}%
\bibitem [{\citenamefont {{Zhu}}\ \emph {et~al.}(2019)\citenamefont {{Zhu}},
  \citenamefont {{Rey}},\ and\ \citenamefont {{Schachenmayer}}}]{spintwa3}%
  \BibitemOpen
  \bibfield  {author} {\bibinfo {author} {\bibfnamefont {Bihui}\ \bibnamefont
  {{Zhu}}}, \bibinfo {author} {\bibfnamefont {Ana~Maria}\ \bibnamefont
  {{Rey}}}, \ and\ \bibinfo {author} {\bibfnamefont {Johannes}\ \bibnamefont
  {{Schachenmayer}}},\ }\bibfield  {title} {\enquote {\bibinfo {title} {{A
  generalized phase space approach for solving quantum spin dynamics}},}\
  }\href@noop {} {\bibfield  {journal} {\bibinfo  {journal} {arXiv e-prints}\
  ,\ \bibinfo {eid} {arXiv:1905.08782}} (\bibinfo {year} {2019})}\BibitemShut
  {NoStop}%
\bibitem [{\citenamefont {Nowak}\ \emph {et~al.}(2011)\citenamefont {Nowak},
  \citenamefont {Sexty},\ and\ \citenamefont {Gasenzer}}]{2011PRB_gasenzer}%
  \BibitemOpen
  \bibfield  {author} {\bibinfo {author} {\bibfnamefont {Boris}\ \bibnamefont
  {Nowak}}, \bibinfo {author} {\bibfnamefont {D\'enes}\ \bibnamefont {Sexty}},
  \ and\ \bibinfo {author} {\bibfnamefont {Thomas}\ \bibnamefont {Gasenzer}},\
  }\bibfield  {title} {\enquote {\bibinfo {title} {Superfluid turbulence:
  Nonthermal fixed point in an ultracold bose gas},}\ }\href {\doibase
  10.1103/PhysRevB.84.020506} {\bibfield  {journal} {\bibinfo  {journal} {Phys.
  Rev. B}\ }\textbf {\bibinfo {volume} {84}},\ \bibinfo {pages} {020506}
  (\bibinfo {year} {2011})}\BibitemShut {NoStop}%
\bibitem [{\citenamefont {Nowak}\ \emph {et~al.}(2012)\citenamefont {Nowak},
  \citenamefont {Schole}, \citenamefont {Sexty},\ and\ \citenamefont
  {Gasenzer}}]{2012PRA_gasenzer}%
  \BibitemOpen
  \bibfield  {author} {\bibinfo {author} {\bibfnamefont {Boris}\ \bibnamefont
  {Nowak}}, \bibinfo {author} {\bibfnamefont {Jan}\ \bibnamefont {Schole}},
  \bibinfo {author} {\bibfnamefont {D\'enes}\ \bibnamefont {Sexty}}, \ and\
  \bibinfo {author} {\bibfnamefont {Thomas}\ \bibnamefont {Gasenzer}},\
  }\bibfield  {title} {\enquote {\bibinfo {title} {Nonthermal fixed points,
  vortex statistics, and superfluid turbulence in an ultracold bose gas},}\
  }\href {\doibase 10.1103/PhysRevA.85.043627} {\bibfield  {journal} {\bibinfo
  {journal} {Phys. Rev. A}\ }\textbf {\bibinfo {volume} {85}},\ \bibinfo
  {pages} {043627} (\bibinfo {year} {2012})}\BibitemShut {NoStop}%
\bibitem [{\citenamefont {Duan}\ \emph {et~al.}(2003)\citenamefont {Duan},
  \citenamefont {Demler},\ and\ \citenamefont {Lukin}}]{2003amospinexchange}%
  \BibitemOpen
  \bibfield  {author} {\bibinfo {author} {\bibfnamefont {L.-M.}\ \bibnamefont
  {Duan}}, \bibinfo {author} {\bibfnamefont {E.}~\bibnamefont {Demler}}, \ and\
  \bibinfo {author} {\bibfnamefont {M.~D.}\ \bibnamefont {Lukin}},\ }\bibfield
  {title} {\enquote {\bibinfo {title} {Controlling spin exchange interactions
  of ultracold atoms in optical lattices},}\ }\href {\doibase
  10.1103/PhysRevLett.91.090402} {\bibfield  {journal} {\bibinfo  {journal}
  {Phys. Rev. Lett.}\ }\textbf {\bibinfo {volume} {91}},\ \bibinfo {pages}
  {090402} (\bibinfo {year} {2003})}\BibitemShut {NoStop}%
\bibitem [{\citenamefont {Davis}\ \emph {et~al.}(2019)\citenamefont {Davis},
  \citenamefont {Bentsen}, \citenamefont {Homeier}, \citenamefont {Li},\ and\
  \citenamefont {Schleier-Smith}}]{2019amospin}%
  \BibitemOpen
  \bibfield  {author} {\bibinfo {author} {\bibfnamefont {Emily~J.}\
  \bibnamefont {Davis}}, \bibinfo {author} {\bibfnamefont {Gregory}\
  \bibnamefont {Bentsen}}, \bibinfo {author} {\bibfnamefont {Lukas}\
  \bibnamefont {Homeier}}, \bibinfo {author} {\bibfnamefont {Tracy}\
  \bibnamefont {Li}}, \ and\ \bibinfo {author} {\bibfnamefont {Monika~H.}\
  \bibnamefont {Schleier-Smith}},\ }\bibfield  {title} {\enquote {\bibinfo
  {title} {Photon-mediated spin-exchange dynamics of spin-1 atoms},}\ }\href
  {\doibase 10.1103/PhysRevLett.122.010405} {\bibfield  {journal} {\bibinfo
  {journal} {Phys. Rev. Lett.}\ }\textbf {\bibinfo {volume} {122}},\ \bibinfo
  {pages} {010405} (\bibinfo {year} {2019})}\BibitemShut {NoStop}%
\end{thebibliography}
%

\end{document}